\documentclass[fleqn,usenatbib]{mnras}

\usepackage[T1]{fontenc}

\usepackage{graphicx}	% Including figure files
\usepackage{amsmath}	% Advanced maths commands
\usepackage{amssymb}	% Extra maths symbols
\usepackage{tabularx}
\usepackage{booktabs}
\usepackage{multirow}
\usepackage{newtxtext,newtxmath}
\usepackage{comment}
\title[X-ray spectroscopy of NGC 55 ULX-1]{Unveiling the disc structure in ultraluminous X-ray source NGC 55 ULX-1}

\author[F. Barra et al.]{
F. Barra,$^{1,2}$\thanks{E-mail: francesco.barra@unipa.it}
C. Pinto,$^{2}$\thanks{E-mail: ciro.pinto@inaf.it}
D. J. Walton,$^{3,4}$
P. Kosec,$^{5}$
A. D'Aì,$^{2}$
T. Di Salvo,$^{1}$
M. Del Santo,$^{2}$
H. Earnshaw, $^{6}$
\and
\ A. C. Fabian,$^{4}$
F. Fuerst,$^{7}$
A. Marino,$^{1,2,8,9}$
F. Pintore,$^{2,10}$
A. Robba,$^{1,2}$
T. P. Roberts.$^{11}$
\\
\\
$^{1}$Universit\`a degli Studi di Palermo, Dipartimento di Fisica e Chimica, via Archirafi 36, I-90123 Palermo, Italy\\
$^{2}$INAF/IASF Palermo, via Ugo La Malfa 153, I-90146 Palermo, Italy\\
$^{3}$Centre for Astrophysics Research, University of Hertfordshire, College Lane, Hatfield AL10 9AB, UK\\
$^{4}$Institute of Astronomy, University of Cambridge, Madingley Road, Cambridge CB3 0HA, UK\\
$^{5}$MIT Kavli Institute for Astrophysics and Space Research, Cambridge, MA 02139, USA\\
$^{6}$Cahill Center for Astronomy and Astrophysics, California Institute of Technology, Pasadena, CA 91125, USA\\
$^{7}$Quasar Science Resources S.L for European Space Agency (ESA), European Space Astronomy Centre (ESAC), Camino Bajo del Castillo s/n, 28692 \\ \ Villanueva de la Ca˜nada,  Madrid, Spain\\
$^{8}$Institute of Space Sciences (ICE, CSIC), Campus UAB, Carrer de Can Magrans s/n, E-08193 Barcelona, Spain\\
$^{9}$Institut d'Estudis Espacials de Catalunya (IEEC), E-08034 Barcelona, Spain\\
$^{10}$INAF - IASF Milano, via E. Bassini 15, I-20133 Milano, Italy\\
$^{11}$Centre for Extragalactic Astronomy \& Department of Physics, Durham University, Department of Physics, South Road, Durham DH1 3LE, UK\\
}

\date{Accepted 2022 August 25. Received 2022 June 08; in original form 2022 March 24}

\pubyear{2022}

\begin{document}
\label{firstpage}
\pagerange{\pageref{firstpage}--\pageref{lastpage}}
\maketitle

\begin{abstract}    % 200 words for Letters
Ultraluminous X-ray sources (ULXs) are the most extreme among X-ray binaries in which the compact object, a neutron star or a black hole, accretes matter from the companion star and exceeds a luminosity of $10^{39} \ \rm erg \ s^{-1}$ in the X-ray energy band alone.
%%% An important class of these object are the ultraluminous supersoft sources (ULSs), that are characterized by their supersoft spectra dominated by a cool blackbody component with kT $\sim 0.1 \ \rm keV$ and low emission above 1 keV. This class of object, initially, were though as a different class of sources respect to the classical ULX but the recent discovery of winds in archetypal ULXs and variable hard tails in some ULSs suggested that the latter are similar to super Eddington accretors but viewed at different angle with respect to the disc-wind cone and with the presence of a thicker wind along the LOS, blocking and reprocessing disc X-ray photons. 
Despite two decades of studies, it is still not clear whether ULX spectral transitions are due to stochastic variability in the wind or variations in the accretion rate or in the source geometry. The compact object is also unknown for most ULXs. In order to place constraints onto such scenarios and on the structure of the accretion disc, we studied the temporal evolution of the spectral components of the variable source NGC 55 ULX-1. Using recent and archival data obtained with the XMM-\textit{Newton} satellite, we modelled the spectra with two blackbody components {which we interpret as thermal emission from the inner accretion flow and the regions around or beyond the spherization radius}. The luminosity-temperature (L-T) relation of each spectral component agrees with the L$ \ \propto$ T$^{4}$ relationship expected from a thin disc model, which suggests that the accretion rate is close to the Eddington limit. However, there are some small deviations at the highest luminosities, possibly due to an expansion of the disc and a contribution from the wind at higher accretion rates. Assuming that such deviations are due to the crossing of the Eddington or supercritical accretion rate, we estimate a compact object mass of 6-14$ M\textsubscript{\(\odot\)}$, favouring a stellar-mass black hole as the accretor.
\end{abstract}


%%%%%%%%%%%%%%%%%%%%%%%%%%%%%%%%%%%%%%%%%%%%%%%%%%

%%%%%%%%%%%%%%%%% BODY OF PAPER %%%%%%%%%%%%%%%%%%

\section{Introduction}
\label{Sec:Introduction}

%%%\textcolor{blue}{Something about ULXs, their main characteristics, super-Eddington accretion, the soft ULXs, open questions, the source studied in this work (previous results), structure of the paper.} \\

Ultraluminous X-ray sources (ULXs) are empirically defined as bright, non-nuclear, point-like, {mainly} extragalactic sources with an X-ray luminosity $ \rm L_X > 10^{39} \, \rm erg \, s^{-1}$ (\citealt{Kaaret_2017}). They are brighter than the Eddington luminosity limit for a $10 M\textsubscript{\(\odot\)}$ black hole (BH) and {can even reach luminosities in excess of}  $10^{41} \, \rm erg \, s^{-1}$ in the canonical X-ray band (0.3-10 keV, \citealt{Walton_2021}).

%%%The first resolved image that allowed to discover ULXs was obtained in 1981 by the Einstein X-ray observatory.
%%%These sources are binary systems composed of a compact object, that can be a NS or a BH, that accretes matter from a companion star. As all X-ray binaries, ULXs are characterised by the mass and spin of the compact object, the mass of its companion star, the orbital separation and the eccentricity. These are responsible for their observed multi-wavelength emission (from hard X-rays band to radio, peaking in the X-rays below 10 keV). The spectral shape and timing behavior of a ULX ultimately depend on the matter transfer rate, the disc geometry, and their variability. Owing to their luminosities, it is clear that ULXs may allow us to understand and investigate the accretion process at extremely high transfer rates.
%%%These objects show a different behavior depending on the characteristics discussed above. Some ULXs, in particular, exhibit a strong X-ray variability at short timescales (minutes, sometimes associated to winds crossing the line of sight) or at long timescales (from days to years, likely due to orbital or super-orbital motions). These confirm the presence in ULXs of an accreting compact object whose mass will turn out to be very difficult to measure.
%%%\textcolor{red}{SUPER EDDINGTON ACCRETORS}. 
Many conjectures have been made to account for the high luminosity of these sources. Initially, ULXs were thought to be powered by black holes with masses greater than 10\,$M\textsubscript{\(\odot\)}$ and, potentially, in the intermediate-mass regime ($10^{\,2-5}\,M\textsubscript{\(\odot\)}$, \citealt{Miller_2004}) {with the best IMBH candidate being HLX-1} (\citealt{Webb2012}). An alternative scenario suggested stellar-mass black holes whose light was beamed along the line of sight (LOS) of the observer by a thick disc-wind cone (e.g. \citealt{King_2001, Poutanen_2007}). However, the discovery of coherent X-ray pulsations from M\,82 ULX-2 with NuSTAR provided unambiguous evidence in support of a ULX hosting a neutron star (NS, \citealt{Bachetti_2014}). Other notable examples are NGC 7793 P13 and NGC 5907 ULX-1 with similar properties (\citealt{Fuerst_2016,Israel_2017a}). 
%%% The source M82 X-2, a transient X-Ray source located in the core of the starburst M82 galaxy, generated a signal characterized by coherent pulsations, providing a clear signature of a spinning NS and ruling out the presence of a black hole. 
NGC 7793 P13, the second discovered pulsating ULX (or pULX), is also the only ULX for which an upper limit to the mass of the compact object was dynamically obtained ($15 M\textsubscript{\(\odot\)}$, \citealt{Motch2014}). The pulsating source NGC 5907 ULX-1 reaches $L_{X} \sim 10^{41} \, \rm erg \, s^{-1}$ (\citealt{Fuerst_2017}), which corresponds to 500 times the Eddington limit of a NS, \textit{de facto} making it the brightest NS known. As of today, only 10 pulsating ULXs (including transient ULXs, see e.g. \citealt{King2020} and references therein) are confirmed among the {$\sim$ 1800} ULXs known (\citealt{Walton_2021}). However, the fraction of pulsating neutron stars might be higher in ULXs as only about 30 ULXs have data with sufficiently good quality to search for pulsations. This suggests that neutron stars might power $\gtrsim 30$\,\% of the nearby, bright, ULXs (see e.g. \citealt{Rodriguez_2020}).

\begin{figure}%[H]
		\centering
		\includegraphics[width=0.48\textwidth]{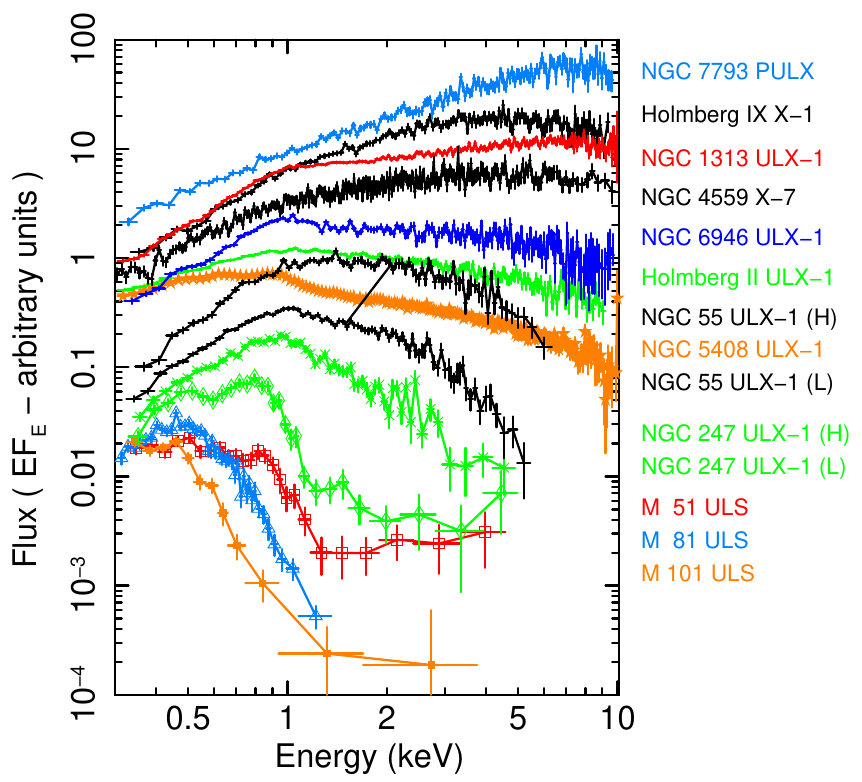}
        \vspace{-0.5cm}
		\caption{{\small X-ray spectra of some brightest ULXs with the hardness increasing from the bottom to top. Note how the high- and low-flux spectra of NGC 55 ULX-1 link the ULX spectra with soft and intermediate hardness. Adapted from \citet{Pinto_2017}. }} 
		%%%  Adapted from \citet{Pinto_2020a}.\protect
		\label{fig: Comparison X-ray spectra of the brightest ULX known}
		\end{figure}

The so-called \textit{ultraluminous state} \citep{Gladstone_2009} is characterized by a strong curvature between 2--10\, keV (e.g. \citealt{Walton_2020} and references therein), and often a soft excess below 2 keV (see Fig. \ref{fig: Comparison X-ray spectra of the brightest ULX known}). \citet{Sutton_2013} classified ULXs into three main regimes according to their spectral slope in the 0.3--10 keV band (soft ultraluminous/SUL for $\Gamma>2$ or hard ultraluminous/HUL for $\Gamma<2$). In the latter case, if the X-ray spectrum has a single peak and is dominated by a blackbody-like component in the 2--5 keV band, it is classified as in the broadened disc (BD) regime. ULXs often switch between different spectral regimes (\citealt{Walton_2020}). The presence of the {low-energy} spectral turnover rules out models of sub-Eddington accretion onto intermediate mass black holes {because the temperature of the soft component would require black hole masses of up to 1000s $M\textsubscript{\(\odot\)}$ with ${\dot M}$ (accretion rate) $\sim 0.01 {\dot M_{Edd}}$, which disagree with their soft spectra.}

Short time-scales variability from seconds to hours has been observed in several ULXs \citep{Heil_2009} {rarely associated to} quasi-periodic patterns \citep{Alston_2021,Strohmayer_2003, Strohmayer_2007,Gurpide_2021a} but is generally at a lower level than in sub-Eddington AGN and black hole X-ray binaries. Long-term variability is also observed on time-scales of a few months and could be associated with super-orbital motions due to precession  and is most common in the HUL regime and, particularly, in pulsating ULXs \citep{Walton_2016,Fuerst_2018,Brightman_2019,Gurpide_2021a}.
\citet{Heil_2010} and \citet{DeMarco_2013} showed that the soft energy band lagged the hard band in the ULX NGC 5408 ULX-1 at frequencies of $\sim$10 mHz.
More recently, very long soft lags of the order of a few ks were found in NGC 55 ULX-1, NGC 1313 ULX-1 and NGC 7456 ULX-1 which show a broad range of spectral hardness (\citealt{Pinto_2017, Kara_2020, Pintore_2020}). The time lags range between a few seconds to ks and it is not clear whether they are produced by the same mechanism.{ They could be produced by down scattering of hard X-ray photons through a thick disc wind cone in the LOS. 
In fact, super-Eddington accretion predicts the launch of powerful, relativistic ($\sim 0.1c$), winds (\citealt{Takeuchi2013}).}

Evidence in support of winds in ULXs was found through the presence of strong, although unresolved, features at soft X-ray energies ($<$ 2 keV) in CCD low-resolution spectra \citep{Strohmayer_2007}. Their time variability and correlation with the source spectral hardness suggested that they might be produced by the ULX itself in the form of a wind (\citealt{Middleton2015b}). This scenario was unambiguously confirmed through the first detection and identification of emission and absorption lines in the high-resolution XMM-\textit{Newton} (here and after XMM) / RGS spectra of NGC 1313 ULX-1 and NGC 5408 ULX-1 (\citealt{Pinto_2016}). In particular, the line-emitting gas is generally close to rest (see, e.g., \citealt{Kosec_2021}) with some exceptions such as NGC 5204 ULX-1 ($v_{\rm LOS}\sim0.3  c$, \citealt{Kosec_2018a}) and in NGC 55 ULX-1 ($0.01-0.08 c$, \citealt{Pinto_2017}). The absorption lines are highly blueshifted  ($0.1-0.3 c$). The ionisation state increases with the outflow velocity and the source spectral hardness, which indicates a detection of hotter and faster phases coming from the inner region at lower inclinations (\citealt{Pinto_2020a}). 

An interesting sub-class of ULXs are ultraluminous supersoft sources (ULSs or SSUL regime). These objects are identified with X-ray spectra dominated by a cool blackbody-like component with $kT \sim 0.1 \ \rm keV$, a bolometric luminosity $\gtrsim 10^{39} \ \rm erg \ s^{-1} $, and very little emission above 1 keV. The discovery of similar winds in archetypal, persistent, ULXs (\citealt{Pinto_2016}) and very recently in NGC 247 ULX-ULS 1 (\citealt{Pinto_2021}), and the presence of a variable hard X-ray tail in some ULSs suggested that they are similar super-Eddington accretors but viewed at a different angle with respect to the disc-wind cone. The fainter hard tail and strong variability such as the presence of dips in the lightcurves of ULSs may be the consequence of a higher inclination angle or a higher accretion rate and a thicker (and variable) wind in the LOS (see, e.g., \citealt{Urquhart2016,Pinto_2017,Alston_2021,D'Ai_2021}).
\\
\\
Despite two decades of dedicated studies several open questions remain unanswered. 
\textit{Are ULX spectral transitions driven by stochastic changes in the wind or variations in the accretion rate / geometry? What is the fraction of matter lost into the wind and, therefore, the net accretion rate onto the compact object? What is the fraction of NS-powered ULXs?}
ULXs which exhibit strong spectral variability are the ideal targets to tackle them. 

\subsection{NGC 55 ULX-1}
ULX-1 is the brightest X-ray source in the NGC 55 galaxy (see Fig.\,\ref{fig:NGC 55 composite}). At a distance of 1.94 Mpc\footnote{https://ned.ipac.caltech.edu}, this source has an X-ray luminosity peak of about $4 \times 10^{39} \ \rm erg \ s^{-1}$ (see, e.g., \citealt{Gurpide_2021a}). The X-ray light curve exhibits sharp drops and 100s-long dips, during which the source flux is quenched in the 2.0 - 4.5 keV band (\citealt{Stobbart_2004}). The spectrum is very soft (if modelled with a powerlaw it yields a slope $\Gamma = 4$, \citealt{Pinto_2017}) and similar to the brightest ULSs, but with a stronger hard tail above 1 keV. It is possible to see from Fig. \ref{fig: Comparison X-ray spectra of the brightest ULX known} that the X-ray spectrum of NGC 55 ULX-1 fits just in between the spectra of bright ULSs and the soft-intermediate spectra of ULXs and, therefore, the source can be considered as a link between these subclasses of ULXs.
%%%It is now believed that ULSs are ULXs in which the wind is geometrically thicker, as said before, thereby highly obscuring the innermost regions emitting hard photons (above 1 keV), as a result of extremely high mass accretion rate or high viewing angles. This was confirmed by the detections of winds and a hard X-ray tail in some ULSs. What is currently missing is a detailed broadband study aimed at understanding the ULX transitions in this low-hardness end between soft and supersoft states. 
NGC 55 ULX-1 is the ideal target for our study as it is very bright, has a spectral curvature above 1 keV similar to (although less severe than) ULSs and both broadband and wind properties are halfway between ULSs and soft-intermediate ULXs (see \citealt{Pinto_2020a}).
Moreover, its X-ray luminosity often crosses the  $ 10^{39} \ \rm erg / s $ threshold, giving an opportunity to test disc structural changes around the Eddington limit, {assuming the accretor is a stellar mass BH or a NS}.
Finally, its flux and spectral variability provide the workbench necessary to {fit L-T trends and break model degeneracies}.

%In order to answer to some of these questions a large team of scientists led by C. Pinto has undergone a series of large campaign primarily focused on XMM-\textit{Newton} observations and follow-up with NUSTAR and Chandra. A large campaign (700+ ks) was dedicated to the study of NGC 1313 ULXs, with primary focus on the continuum-wind relationship in NGC 1313 ULX-1 and the search for pulsations in NGC 1313 ULX-2. Pulsations were indeed detected in ULX-2 (\citealt{Sathyaprakash2019a}) and the spectral evolution showed the presence of a wind and an overall behavior driven by the long-term variations in the accretion rate (\citealt{Robba2021}). ULX-1 broadband analysis has shown a complex behavior (\citealt{Walton_2020}) likely due to the presence of a powerful wind along the line-of-sight that varies with the accretion state (\citealt{Pinto_2020b}). Another deep campaign 800+ ks has focused on the supersoft NGC 247 ULX-1 which showed a remarkable dipping behavior characterised by quasi-periodicities (\citealt{Alston_2021}) likely produced by the variable wind in the line of sight (\citealt{Pinto_2021}). However an alternative scenario invoking the propeller effect during periods of lower accretion rate might also explain the source behavior (\citealt{D'Ai_2021}). 

In order to study the spectral variability of NGC 55 ULX-1 and the wind response to changes in the broadband flux, our team has requested and has been awarded with three full XMM orbits in different AOs (about 390 ks, PI: Pinto).
The first observation occurred in 2018 and caught the source in a low state, while the two latter ones were triggered in 2020 and 2021 in order to obtain well-exposed RGS spectra during intermediate-high states. More detail on the triggering and the RGS analysis will be provided in a forthcoming paper.

This paper is the first in a series and will focus on the broadband spectral variability. This paper is organized as follows. In Sect. \ref{Observations and Spectral modelling} we report the details on the observations, the spectral modeling  in Sect. \ref{Spectral modelling}. In Sect. \ref{Discussion} we discuss our results and provide our conclusions in Sect. \ref{Conclusions}. All uncertanties are at 1$\sigma$ (68 \% level).

%%% The primary goals of this proposal was to study if indeed the winds can become optically thick enough to block and reprocess the disc X-ray photons, making the source appear as a soft thermal emitter or ultraluminous supersoft X-ray source.The new and archival data will enable us to perform time- / flux-resolved X-ray spectroscopy of the wind and show how the wind causes the transition from a classical ultraluminous X-ray to a supersoft state. This work will require two independent approach: a careful modelling of the source spectral continuum for each observations (to study the evolution of the accretion rate and source appearance) and a time-resolved study of the wind evolution by means of high-resolution X-ray spectroscopy. This paper focuses on the broadband spectral variability of the source. The high resolution spectroscopy will be presented in a follow up work.

\section{Observations and data reduction}
\label{Observations and Spectral modelling}

%%%\textcolor{blue}{Detail on the observations, the data reduction, lightcurves (Swift and XMM), the Spectral modelling (plots of spectra comparing different observations and some with detailed modelling)}.
XMM-\textit{Newton} observed NGC 55 ULX-1 ten times over a period of 20 years. 
As we can see from Table \ref{table:observations log}, XMM-\textit{Newton} observed NGC 55 ULX-1 six times with reasonably long ($> 30 $ ks) observations, which enabled us to study the evolution of the spectra over two decades. Four additional short ($< 10$ ks net) observations were taken in the recent years. This permits us to probe both short-term (hours-days) and long-term (months-years) variability time scales.

\begin{center}
	\begin{table}%[h]
\caption{Table of observations of the source NGC 55 ULX-1.}  
 \renewcommand{\arraystretch}{1.}
 \small\addtolength{\tabcolsep}{0pt}
 \vspace{0.1cm}
	\centering
	\scalebox{0.9}{%
	\begin{tabular}{ccccccc}
    \toprule
   % \multicolumn{5}{c}{{table model rhbm}} \\
   % \midrule
%   \multicolumn{6}{c}{{Best fit parameters of RHBM model}} \\
    % \midrule
    {{Obs. ID}}  &
    {{Date}}  &
    {{t$\,_{\rm tot}$ [s]}} &
    {{t$\,_{\rm net, \, EPIC-PN}$ [s]}} &
    {{{{CR}$\,_{\rm EPIC-PN}$ [c/s]}}} \\
    
    \midrule
0028740201 & 2001-11-14  & 33619  & 27223  & 1.197 $\pm$ 0.007  \\\midrule
0028740101 & 2001-11-15  & 31518  & 24718  & 0.578 $\pm$ 0.005 \\\midrule
0655050101 & 2010-05-24  & 127437 & 95295  & 0.772 $\pm$ 0.003  \\\midrule
0824570101 & 2018-11-17  & 139800 & 90733  & 0.513 $\pm$ 0.002 \\\midrule
0852610101 & 2019-11-27  & 11000  &  4079  & 1.10 $\pm$ 0.02  \\\midrule
0852610201 & 2019-12-27  &  8000  &  4055  & 1.09 $\pm$ 0.02  \\\midrule
0852610301 & 2020-05-11  &  9000  &  4909  & 0.422 $\pm$ 0.009  \\\midrule
0852610401 & 2020-05-19  &  8000  &  3969  & 0.152 $\pm$ 0.006  \\\midrule
0864810101 & 2020-05-24  & 132800 & 102158 & 0.707 $\pm$ 0.003  \\\midrule
0883960101 & 2021-12-12  & 130200 & 92294  & 0.872 $\pm$ 0.003  \\\midrule
    \bottomrule
    \end{tabular}}\label{table:observations log}
       \vspace{0.3cm}
      \begin{quotation}\footnotesize
      {t$\,_{\rm net}$ is the exposure time after the removal of periods with high solar flares and {CR}$\,_{\rm EPIC-PN}$ is the net source count rate.}
      \end{quotation}
 %    \vspace{-0.3cm}
\end{table}
\end{center}

The data were reduced with the \textit{Science Analysis System} ({\scriptsize{SAS}}) version 18.0.0\footnote{https://www.cosmos.esa.int/web/XMM-\textit{Newton}}.
The raw data were obtained from the XMM-\textit{Newton} Science Archive (XSA)\footnote{https://www.cosmos.esa.int/web/XMM-\textit{Newton}/xsa}. 
We used recent calibration files (February 2021).
%%% We ran the \textit{cifbuild} task to generate the Current Calibration File Index File (CIF). This enables us to retrieve the most recent calibration data necessary for process. Furthermore, it is necessary to update the observational information on the ODF data set with the \textit{odfingest} task. In fact, this task extracts the information from the instruments housekeeping files and from the calibration database and incorporate this information into the ODF summary file producing a file called SUM.SAS. 
We ran the \textit{epproc} and \textit{emproc} tasks to build the EPIC-PN and EPIC-MOS 1,2 event files, respectively. 
These are subsequently filtered for the flaring particle background. We chose the recommended cutting threshold in the lightcurves above 10 keV (count rate $<$ 0.5 c/s for EPIC-PN and $<$ 0.35 c/s for EPIC-MOS 1 and 2).

\begin{figure}%[H]
		\centering
		\includegraphics[width=0.45\textwidth]{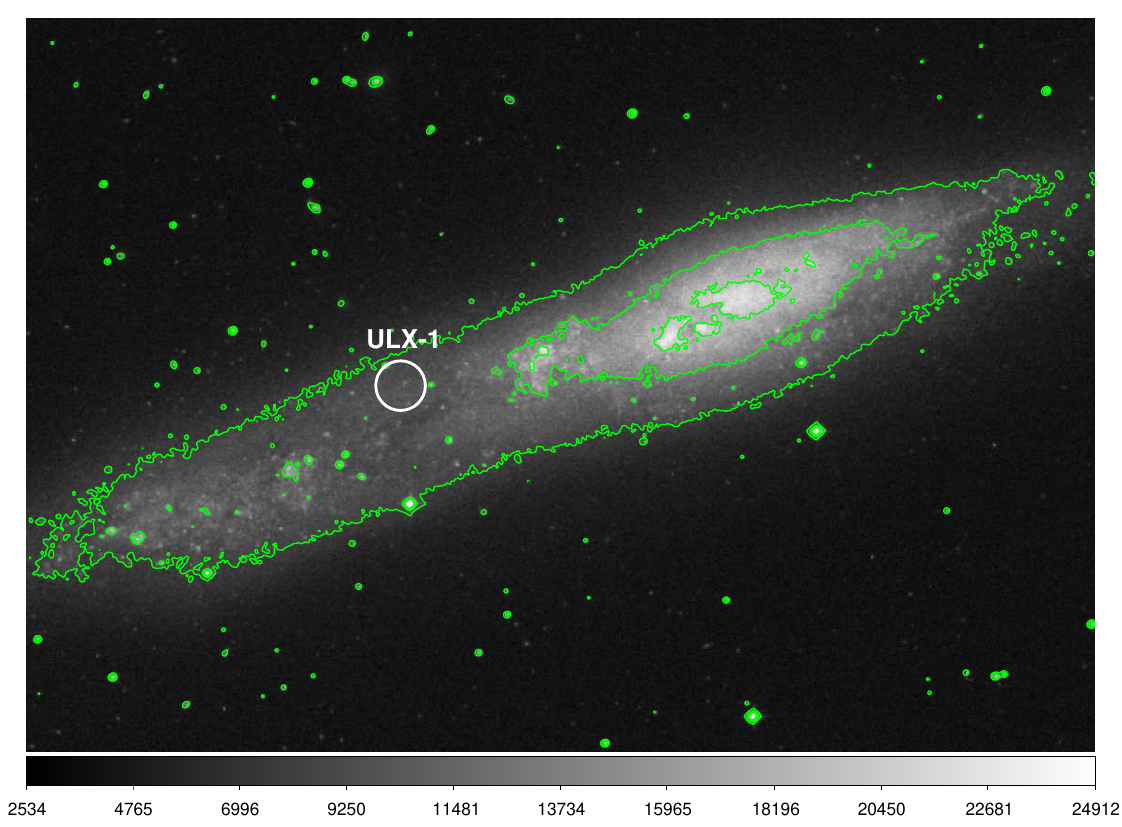}
		\includegraphics[width=0.45\textwidth]{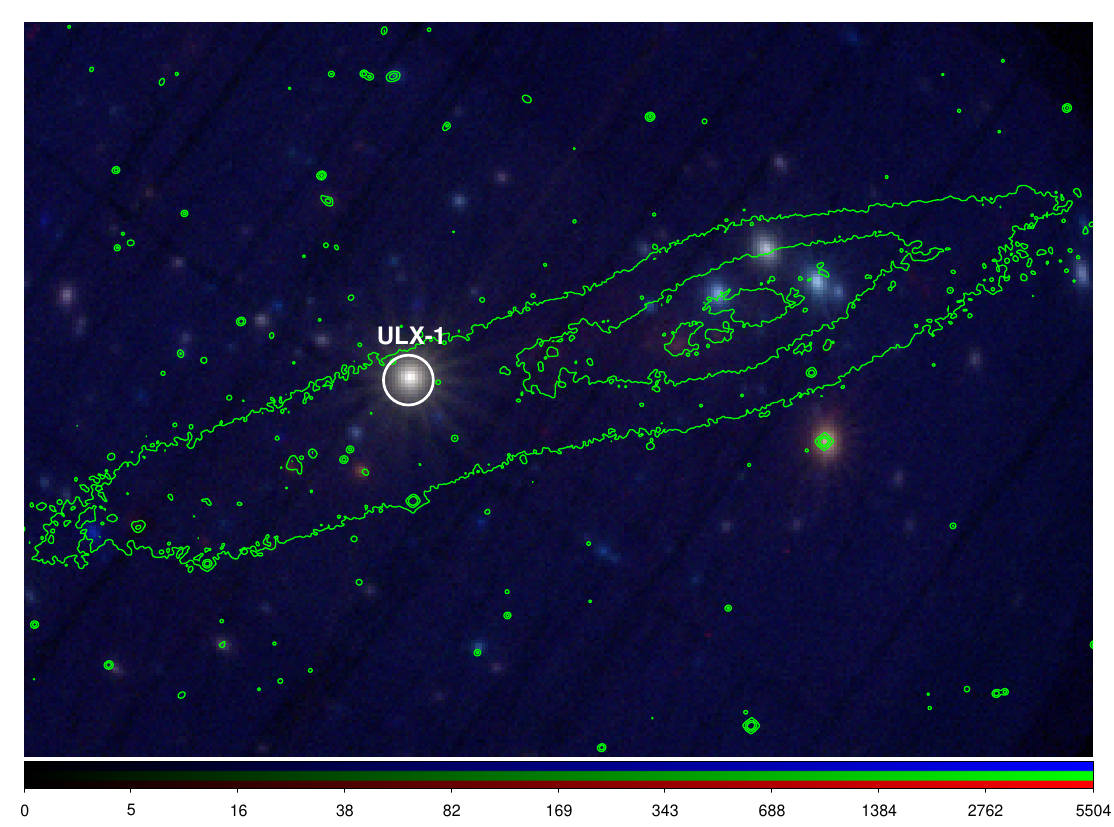}
        % \vspace{-0.2cm}
		\caption{\small Upper panel: Optical image of NGC 55 obtained from the Digitized Sky Survey. 
		\ Lower panel: false-color RGB X-ray image obtained by extracting images in three energy bands (red 0.3-1 keV, green 1-2 keV, blue 2-10 keV) and stacking over all observations. The contours of the optical image are overlaid to show the ULX position in the galaxy.}
		\label{fig:NGC 55 composite}
	\end{figure}

We extracted EPIC MOS 1-2 and PN images in three energy bands to create a false-color RGB image (red 0.3-1 keV, green 1-2 keV, blue 2-10 keV)\footnote{https://sites.google.com/cfa.harvard.edu/saoimageds9}. The images from the same energy band and different observations were stacked to increase the statistics. The final RGB mosaic is shown in Fig.\,\ref{fig:NGC 55 composite} (bottom panel). ULX-1 is the yellow-white, brightest object, in the central-left region. Some fainter and harder (blue) X-ray binaries are present near the galaxy centre. The soft (red) source outside the galaxy contours is a field star in our Galaxy. Overlaid are the contours of surface brightness of the optical image obtained from the Digitized Sky Survey (DSS)\footnote{DSS, https://irsa.ipac.caltech.edu/data/DSS/} (top panel).

\begin{figure*}%[!H]
		\centering
		\includegraphics[height=0.47\textwidth]{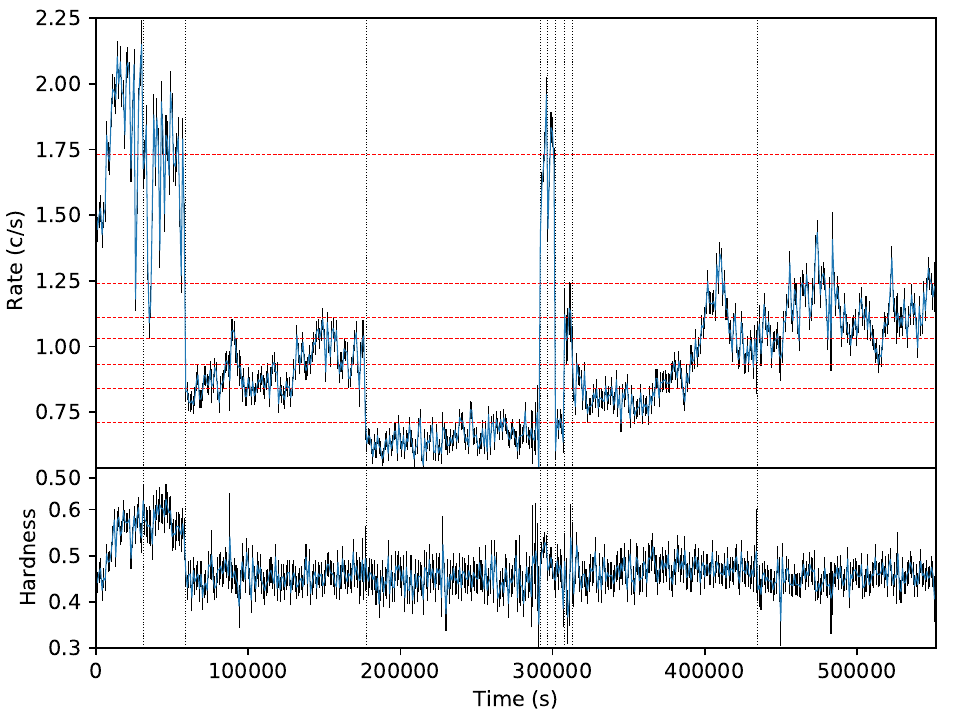}
		\includegraphics[height=0.476\textwidth]{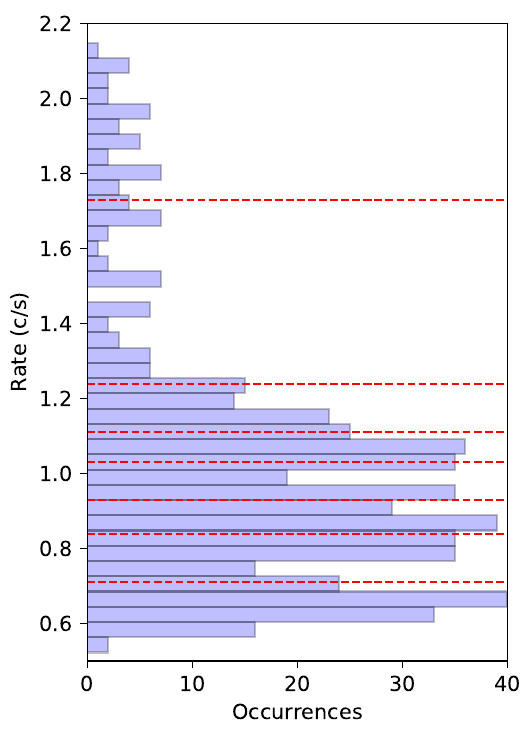} %width=0.32\textwidth,
        \vspace{-0.3cm}
		\caption{{\small Left panel: XMM EPIC-PN lightcurves (top left) of NGC 55 ULX-1 and hardness ratio (bottom left, defined as the 1-10 keV / 0.3-10 keV counts ratio) from 2001 to 2021 {with time bins of 1 ks}. Vertical dotted lines separate the individual XMM observations, which have be attached for displaying purposes. Right panel: count-rate histogram.
		The horizontal red lines indicate the levels chosen to extract eight count-rate resolved spectra.}}
		\label{fig: lightcurves}
\end{figure*}

We also extracted background-corrected lightcurves for ULX-1 and each individual observation after carefully selecting the source and the background regions using the \textit{epiclccorr} task. For the source we selected a circular region of 0.5 arcmin radius centered on {the \textit{Chandra} X-ray estimated position} (RA: 00$^{h}$ 15$^{m}$ 28.89$^{s}$ DEC: -39$^{\circ}$ 13' 18.8''), while for the background we chose a larger circular region a few arc minutes away from the source and avoiding contamination from the copper ring (wherever it was possible) in the same chip of the source region and away from of chip gaps.
There might be a very small loss of counts in the observations, especially for Obs. ID: 0028740101, with the ULX-1 off-axis but this should be marginal given the softness of its spectra. Indeed, despite the larger off-axis angle, its spectrum appears as the brightest and hardest (see Sect. \ref{Sec:spectra}).
% The XMM-\textit{Newton} lightcurves (from a few to 130 ks) enable to investigate the source behavior on much shorter timescales, down to a few hundred seconds depending on the source count rate.
%\begin{figure}%[!htb]
		%\centering
		%\includegraphics[width=0.45\textwidth]{Images/NGC55_Swift_XRT.png}
        %\vspace{-0.3cm}
		%\caption{{\small Left panel: long-term Swift/XRT lightcurve of NGC 55 ULX-1 with %the typical count-rate levels for the high (solid red line) and low (dashed blue %line) states observed with XMM..}}
%		\label{fig: lightcurves}
%\end{figure} 
The 0.3-10\,keV EPIC-PN lightcurves of the individual observations were joined with {{\scriptsize PYTHON}} and shown in Fig. \ref{fig: lightcurves} (top-left panel). The XMM lightcurves confirm that the source flux changed dramatically as anticipated in Sect. \ref{Sec:Introduction}. In the early XMM observations - when the source was brighter - the lightcurve exhibited flux dips lasting hundreds of seconds where its flux dropped by a factor of 2. In the top-right panel we also show the histogram of the count rate for the all-time lightcurve.
%%%, likely due to wind clumps crossing the LOS (\citealt{Stobbart_2004,Pinto_2017}). A lower density in the wind could explain the spectral hardening when the source brightens as for the supersoft NGC 247 ULX-1 (\citealt{Pinto_2021}). 
In particular, defining the hardness ratio (HR) as the ratio between the counts in the 1-10\,keV and the 0.3-10\,keV energy band, respectively, we show that the HR is generally higher when the source is brighter and that
decreases in proximity of the dips (Fig. \ref{fig: lightcurves} bottom-left panel).

\begin{figure}%[!H]
		\centering
		\includegraphics[width=0.45\textwidth]{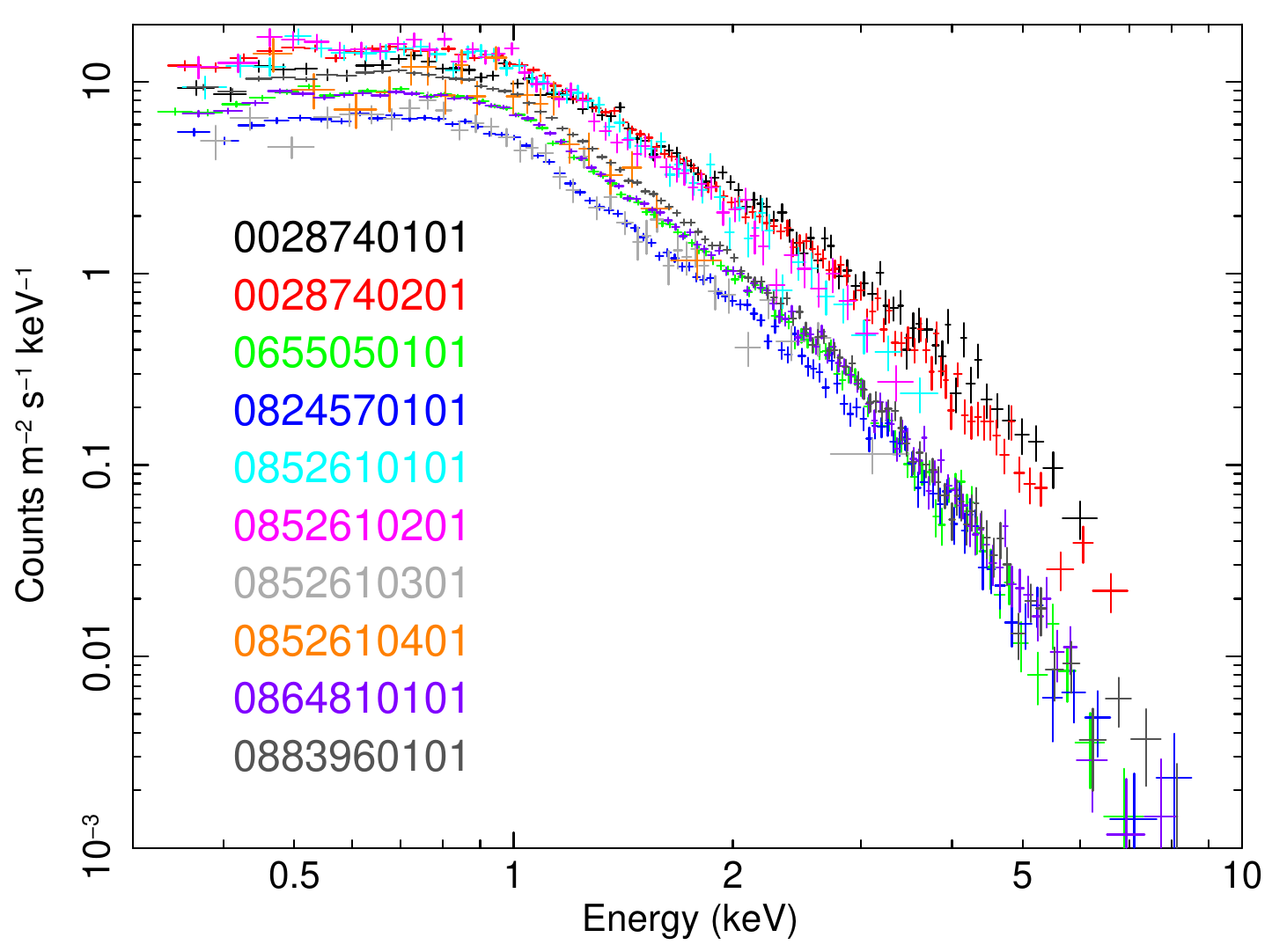}
        \vspace{-0.3cm}
		\caption{{\small EPIC-PN spectra for all observations.}}
		\label{fig: EPIC SPECTRA}
		\includegraphics[width=0.45\textwidth]{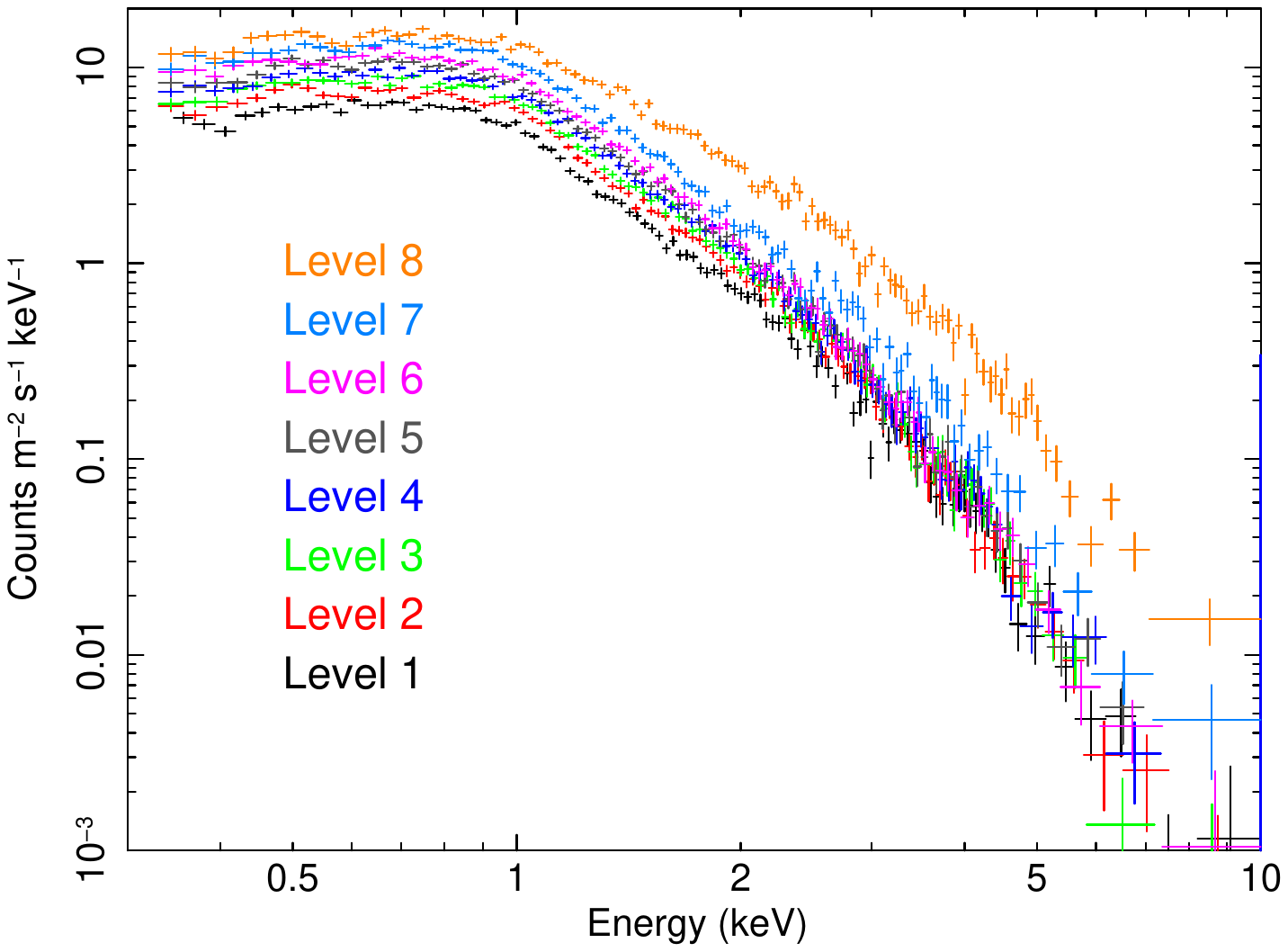}
        \vspace{-0.3cm}
		\caption{{\small EPIC-PN spectra extracted for the eight flux-resolved interval.}}
		\label{fig: EPIC SPECTRA FRS}
	\end{figure}

%This variability of the source can be seen in more detail analysing and modeling the EPIC spectra of the 9 observations in the next sections, after discussing the various spectral models that I use with the software SPEX.
We extracted EPIC PN and MOS spectra for each observations using the same source and background regions chosen for the lightcurves extraction. The \textit{rmfgen} and \textit{arfgen} tasks were used to generate response matrices and effective area files. The EPIC-PN spectra of all the observations are shown in Fig. \ref{fig: EPIC SPECTRA}.

The individual observations have exposure times which differ by up to an order of magnitude and do not show dramatic changes in the spectral hardness despite their substantial flux variability in agreement with the lightcurve. The spectra extracted for some observations are also nearly superimposable (see Fig. \ref{fig: EPIC SPECTRA}). In order to compare spectra with similar statistics, we also extracted spectra in ranges of count rate selected ad-hoc according to the count rate histogram (see also \citealt{Pinto_2017}). We split the XMM all-time lightcurve in eight regimes of count rate as shown in Fig. \ref{fig: lightcurves}. These regimes were chosen in order to balance the total counts for each level (see Table \ref{frs table} for more detail) and some spikes appearing in the count-rate histogram (Fig. \ref{fig: lightcurves} top-right panel). The count-rate selected spectra were then stacked among the different observations in order to obtain eight time-averaged flux-resolved spectra (one per EPIC camera). This selection criterium allow us to probe variability mechanisms at different time scales. The flux-selected EPIC-PN spectra are plotted in Fig. \ref{fig: EPIC SPECTRA FRS}. 

%%% In section \ref{Discussion} we will discuss the main results of our work with particular focus on the spectral evolution as compared to relevant disc models found in the literature. Within this framework we will also provide our conclusions and shed lights on future perspectives.
%%% with an accretion column $N_{H}$ fixed to 2.5 $\cdot$ 10$^{21}$/cm$^{2}$

	\begin{center}
	\begin{table}%[h]
\caption{Total counts and net exposure times of EPIC flux-selected spectra. }  
 \renewcommand{\arraystretch}{1.}
 \small\addtolength{\tabcolsep}{-2pt}
 \vspace{0.1cm}
	\centering
	\scalebox{1}{%
	\begin{tabular}{ccccccc}
    \toprule
   % \multicolumn{5}{c}{{table model rhbm}} \\
   % \midrule
%   \multicolumn{6}{c}{{Best fit parameters of RHBM model}} \\
    % \midrule
    {{ {Range}}}  &
    {{EPIC-PN}}  &
    {{MOS1}}  & {{MOS2}} &
    {{EPIC-PN}} & {{MOS1}} & {{MOS2}} \\
    {{[c/s]}} & {{Cnts}} & {{Cnts}} & {{Cnts}} & {{Exp. [s]}} & {{Exp. [s]}} & {{Exp. [s]}} \\
    
\midrule
       0.50-0.71 & 38392 & 13274 & 13121 & 76795 & 95051 & 95322    \\\midrule
       0.71-0.84  & 43729 & 13977 & 14160 & 72282 & 84108 & 84206    \\\midrule
       0.84-0.93 & 43845 & 14728 & 14699 & 64371 & 79211 & 78945  \\\midrule
       0.93-1.03 & 41879 & 14215 & 13847 & 55327 & 66468 & 66286  \\\midrule
       1.03-1.11 & 42008 & 13816 & 13925 & 50550 & 61191 & 60917  \\\midrule
       1.11-1.24 & 40366 & 12964 & 13021 & 44413 & 53436 & 53513 \\\midrule
       1.24-1.73 & 31926 & 10640 & 10737 & 36211 & 42301 & 42391
       \\\midrule
       1.73-2.25 & 29716 & 10735 & 10657 & 28954 & 32309 & 32332
       \\\midrule
    \bottomrule
    \label{frs table}
    \end{tabular}}
\end{table}
\end{center}

% \subsection{Swift long-term monitoring}
% In order to put the XMM-\textit{Newton} observations in a broader context of source behaviour, we extracted the Swift XRT lightcurve using all the data available by October 2021. Through an online tool of the Swift website (https://www.swift.ac.uk/user\_objects/, \citealt{Evans2009}) we produced Swift/XRT lightcurve for NGC 55 ULX-1. The monitoring covers almost 8 years although is evenly sampled. The XRT lightcurve shows a strong variability by up to a factor 4 on timescales of a few weeks (see \ref{fig: lightcurves} left panel). There is no evidence of periodicities, which suggests that some of it might be due to the stochastic variability of the wind with thick clumps crossing our line of sight (LOS) towards the X-ray source (i.e. the inner accretion disc).

% \begin{figure}%[H]
% 		\centering
% 		\includegraphics[width=0.50\textwidth]{Images/NGC55_Swift_XRT.pdf}
%         \vspace{-0.3cm}
% 		\caption{{\small Long-term Swift/XRT lightcurve of NGC 55 ULX-1 with the typical count-rate levels for the high (solid red line) and low (dashed blue line) states observed with XMM.}}
% 		\label{fig: SWIFT/XRT lightcurves}
% \end{figure} 

% % NGC 55 Swift lc
%   \begin{figure}%[h!]
%   \centering
% %% \includegraphics[width=6cm]{Images/immagine_rgb.png}
%   \includegraphics[width=8cm]{Images/NGC55_Swift_XRT.pdf}
%       \caption{{\swift} long-term lc ....}
%     \label{image_rgb}
%   \end{figure}

\section{Spectral modelling}
\label{Spectral modelling}

The spectra were modelled with the {\scriptsize{SPEX}} fitting package v3.06 \citep{Kaastra_1996}. In order to use $\chi^{2}$ statistics, we rebinned the EPIC PN and MOS 1,2 spectra with bins of at least 1/3 of the spectral resolution and with at least 25 counts using the {\scriptsize{SAS}} task \textit{specgroup}. All emission components were corrected for absorption from the circumstellar and interstellar medium with the \textit{hot} model (freezing the temperature of the gas to $10^{-4}$ keV, which provides a neutral gas in {\scriptsize{SPEX}}). For all emitting and absorbing plasma components we adopted the recommended Solar abundances of \citet{Lodders2009} which are the default abundances in {\scriptsize{SPEX}}. The spectral models accounted for the source redshift as well ($z=0.00043$\footnote{https://ned.ipac.caltech.edu}). Each spectral model is fitted simultaneously to the EPIC MOS 1,2 and PN spectra as they overlap in the 0.3-10 keV energy band. 

\subsection{Testing different spectral models for Obs. ID 0655050101}
It is common to use multiple thermal models (e.g. blackbody emission components) to reproduce ULX spectra (see, e.g., \citealt{Stobbart_2006,Pintore_2015,Walton_2018,Gurpide_2021a}).
At first, we tested several models with the spectra from the observation 0655050101. On the one hand we wanted to compare the new \textsc{SPEX} code used for this spectral modelling with the previous versions and other codes, {e.g. {\scriptsize{XSPEC}} or other {\scriptsize{SPEX}} versions}, used in the recent years for the same source (see, e.g., \citealt{Pinto_2017}, \citealt{Pintore_2015}). On the other hand, the comparison of the $\chi^{2}$ values (and degrees of freedom or d.o.f.) from different spectral models constrains the best fitting continuum model. Finally, we aimed at understanding the systematic effects {when we have not modelled the features.}

The spectral models that we tested were based on different combinations of the following continuum components: blackbody emission (\textit{bb} model in {\scriptsize{SPEX}}), blackbody modified by coherent \footnote{ {\scriptsize{SPEX}} manual for more detail at  https://personal.sron.nl/~jellep/spex/manual.pdf} Compton scattering (\textit{mbb}), multi-temperature disc blackbody (\textit{dbb}), and Comptonization of soft photons in a hot plasma (\textit{comt}). The components used to fit emission and absorption lines are: gaussian line (\textit{gaus}), collisional-ionisation equilibrium emission (\textit{cie}) and photoionisation-equilibrium absorption (\textit{xabs}, see {\scriptsize{SPEX}} manual for more details).

Here there is a summary of the models tested. \\
- {RHB}: simple blackbody emission (B) corrected for redshift (R) and neutral interstellar absorption (H).\\
- {RHBB}: two blackbody components are used to account for a different structure in the inner and outer regions of the accretion disc.\\
- {RHBD}: a cool blackbody describes the outer disc and a disc blackbody (D) reproduces the inner disc.\\
- {RHBM}: a modified blackbody model (M) describes the inner disc.\\
- {RHBCom}: a Comptonization emission is added to the RHB model.\\
- {RHBMC}: a \textit{cie} emission model (C) is added to the RHBM model.\\
- {RHBMCX}: The \textit{xabs} absorption model is added to the RHBMC.\\
- {RHBMCG(G)}: One or more gaussian lines (G) are used to fit emission and absorption lines.

\begin{figure}%[H]
		\centering
		\includegraphics[width=0.45\textwidth]{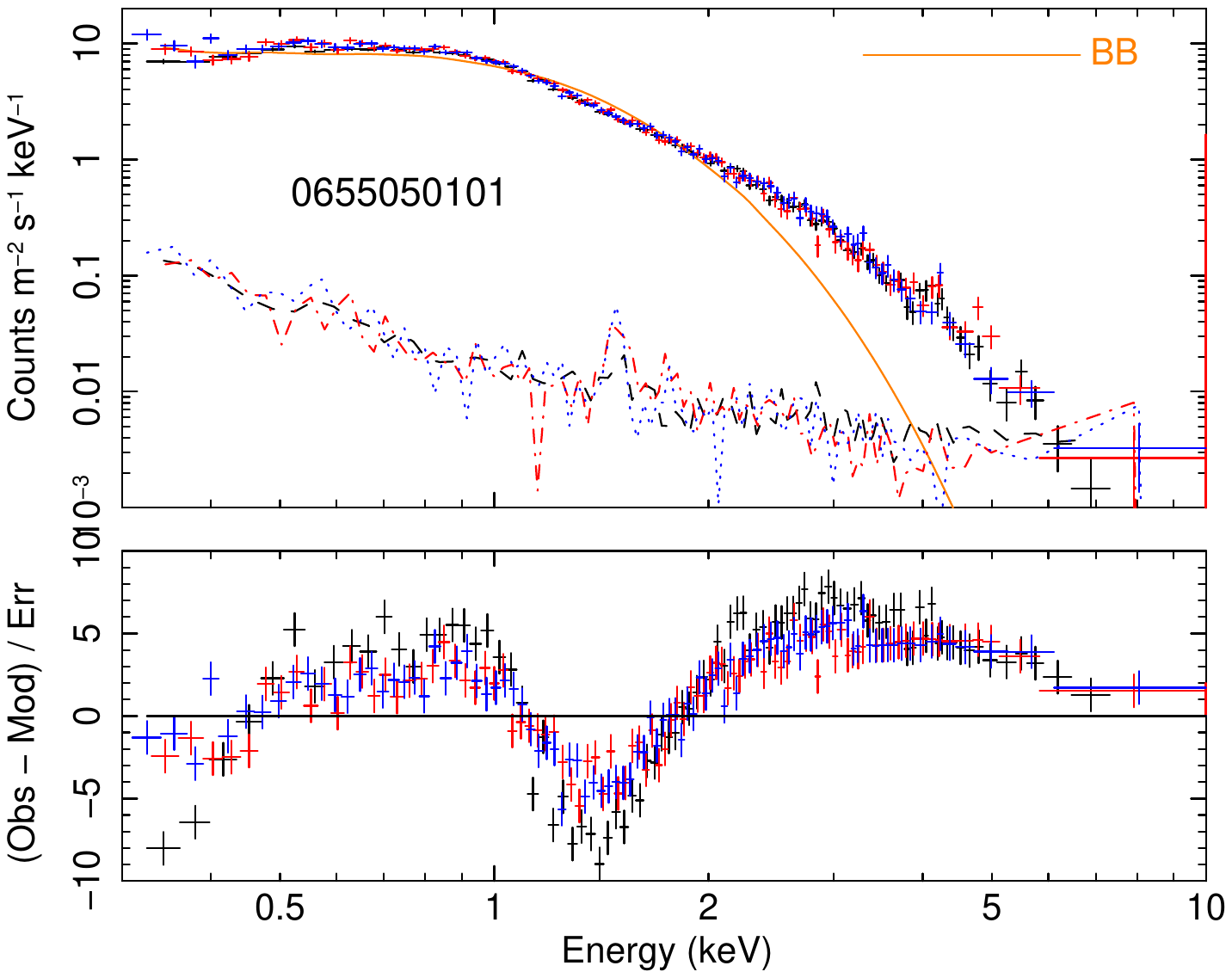}
		\includegraphics[width=0.45\textwidth]{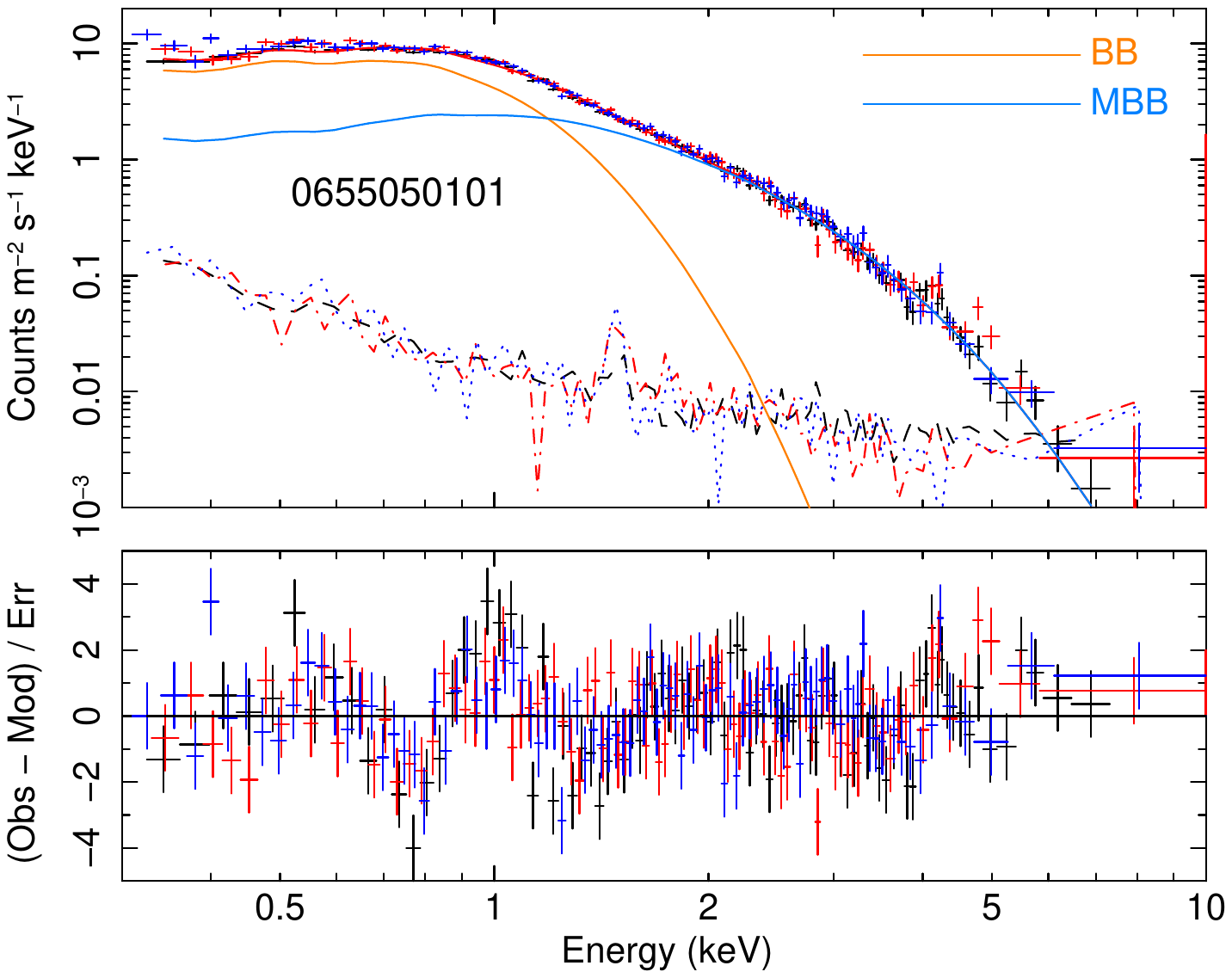}
		\includegraphics[width=0.45\textwidth]{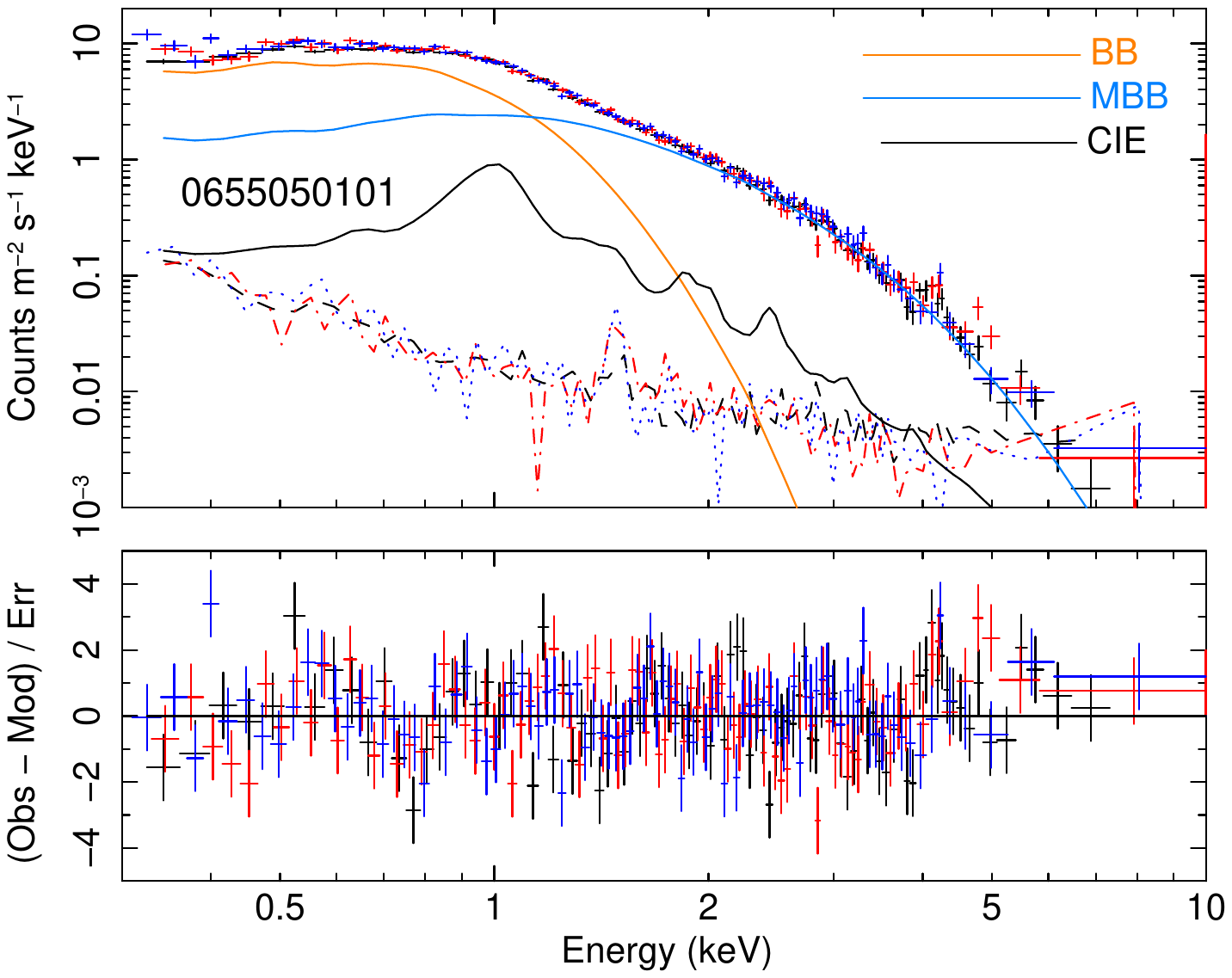}
        \vspace{-0.3cm}
		\caption{{\small NGC 55 ULX-1 EPIC MOS and PN spectral fits performed using the RHB model (top panel), RHBM model (middle panel) and the RHBMC model (bottom panel) for the Obs. ID: 06555050101. {The dashed-dotted lines represent the background spectra.}}}
		\label{fig: Spectral fit for 06555050101 observation using the RHBM model}
		\end{figure}

The first 5 models were primarily used to reproduce the broadband shape and continuum spectrum of the source.  All the continuum models leave strong residuals around 1 keV (positive) and 0.7-0.8 keV and 1.2-1.3 keV (negative), see e.g. Fig. \ref{fig: Spectral fit for 06555050101 observation using the RHBM model}. These were resolved in groups of narrow lines with the high-resolution RGS spectrometers \citep{Pinto_2017,Pinto_2021}.

The other 4 models mentioned above were therefore used to account for any wind features. The \textit{cie} emission model was used mainly to describe the broad emission feature at around 1 keV, which is the most intense. CCD detectors cannot resolve the small velocity shift of the emission lines and we therefore considered the \textit{cie} to be at rest. The dominant absorption component of the wind was found to be outflowing at $0.2c$ (\citealt{Pinto_2017}) which is a shift large enough to be detected even with the EPIC detectors. { However, we chose to fix the outflow velocity for the \textit{xabs} absorption component to this value, in order to avoid model degeneracies with the ionization parameter $\xi$}. Both absorption and emission lines were found to be narrow ($\sigma_{V} \lesssim1000$ km/s) in RGS, which cannot be resolved by EPIC and, therefore, we adopted the default velocity dispersion of 100 km/s for the \textit{cie} and \textit{xabs} components (EPIC resolution $\gg$ 1000 km/s around 1 keV). The gaussian model provided an alternative phenomenological approach to measure the strength of the dominant spectral emission feature around 1 keV and the absorption features around 0.75 keV and 1.25 keV.

In Fig. \ref{fig: Spectral fit for 06555050101 observation using the RHBM model} we show three representative examples of our fits for Obs. ID 06555050101 showing a single \textit{bb} component continuum model (top panel), a \textit{bb+mbb} model (middle panel) and a model which also accounts for wind emission lines (\textit{bb+mbb+cie}, bottom panel). As expected, a single thermal component provided a very poor description of the spectral continuum. Two thermal models were able to fit the broadband shape, while the inclusion of the \textit{cie} accounts for the dominant wind emission feature at 1 keV.

In Table \ref{table: XMM 0655050101 spectral fits} we report the results from our fits of Obs. ID 06555050101. Among all pure-continuum models the best-fit one turned out to be the RHBM model, composed of a cool ($\sim0.16$ keV) blackbody and a hot ($\sim0.7$ keV) blackbody modified by coherent Compton scattering (see {\scriptsize{SPEX}} manual for more details), although comparable results were achieved by a hot disc-blackbody or Comptonization component. The results obtained with the RHBM model on the observation 0655050101 were fully consistent with \citet{Pinto_2017}. We therefore decided to keep such model as the baseline continuum model for the rest of the analysis.

The total column density {of cold gas in the LOS towards the source}, $N_{H}$, was about $2.5 \times 10^{21} \, \rm cm^{-2}$, a few times larger than the Galactic value ($7 \times 10^{20} \, \rm cm^{-2}$)\footnote{https://heasarc.gsfc.nasa.gov/cgi-bin/Tools/w3nh/w3nh.pl}. This suggested that a substantial amount of gas found along the LOS is located in the circumstellar medium around the ULX or in the host galaxy.

The inclusion of the wind (both emission and absorption models) significantly improved the overall quality of the fits. The addition of the wind emission corresponded to a decrease in the overall $\chi^2$ of 73 for 2 additional d.o.f. (normalisation and temperature of the \textit{cie}) which flattened most residuals around 1 keV (see Fig. \ref{fig: Spectral fit for 06555050101 observation using the RHBM model}). The addition of the wind absorption yields a further $\Delta \chi^2 = 18$ for 2 more d.o.f. (column density and ionisation parameter of the \textit{xabs}). 
%&&We will discuss the wind properties in Sect. \ref{Discussion}}. 
However, it is important to 
notice that these additional line components did not strongly affect the continuum parameters nor the total bolometric luminosity as previously found by \citet{Pinto_2020b} and \citet{Walton_2020}.

	\begin{center}
	\begin{table*}%[h]
\caption{Results from the modeling of the XMM-\textit{Newton} spectrum of NGC 55 ULX-1 with the data of the observation 0655050101. }  
 \renewcommand{\arraystretch}{1.}
 \small\addtolength{\tabcolsep}{-4pt}
 \vspace{0.1cm}
	\centering
	\scalebox{1}{%
	\begin{tabular}{ccccccccccc}
    \toprule
   % \multicolumn{5}{c}{{table model rhbm}} \\
   % \midrule
%   \multicolumn{6}{c}{{Best fit parameters of RHBM model}} \\
    % \midrule
    {{Parameter /}}  &
    {{RHB }}  &
    {{RHBB}}  & {{RHBD}} &
    {{RHBM}} & {{RHBCom}} & {{RHBMC}} & {{RHBMCX}} & {{RHBMG}} & {{RHBMGG}} \\
    {{component}} &{{Model}}  &  {{Model}}   & {{Model}}  & {{Model}} & {{Model}} & {{Model}} & {{Model}} & {{Model}} & {{Model}}\\ 
    
    \midrule
       $L_{X\,bb1}$   &  0.72 $\pm$ 0.01 & 0.98 $\pm$ 0.09 &  0.9 $\pm$ 0.1 & 0.93 $\pm$ 0.02  & 0.8 $\pm$ 0.1 & 0.8 $\pm$ 0.1 & 1.0 $\pm$ 0.3 & 0.8 $\pm$ 0.1  & 1.3 $\pm$ 0.3   \\\midrule
       $L_{X\,bb2}$  &  --- \par & 0.36 $\pm$ 0.02 & --- \par  & --- \par & --- \par & --- \par & --- \par & --- \par & --- \par  \\\midrule
       $L_{X\,mbb}$  &  --- \par & --- \par & --- \par & 0.57 $\pm$ 0.02  & --- \par & 0.56 $\pm$ 0.03 & 0.6 $\pm$ 0.1 & 0.61 $\pm$ 0.03 &  0.63 $\pm$ 0.03  \\\midrule
       $L_{X\,dbb}$  &  --- \par & --- \par & 0.58 $\pm$ 0.04  & --- \par  & --- \par & --- \par & --- \par & --- \par & --- \par  \\\midrule
       $L_{X\,comt}$  & --- \par & --- \par & --- \par & --- \par & 0.7 $\pm$ 0.1 & --- \par & --- \par & --- \par & --- \par  \\\midrule 
       $L_{X\,CIE}$  & --- \par & --- \par & --- \par & --- \par & --- \par  & 0.07 $\pm$ 0.01 & 0.08 $\pm$ 0.02 & --- \par & --- \par  \\\midrule 
       $L_{X\,gauss}$  & --- \par & --- \par & --- \par & --- \par &  --- \par & --- \par & --- \par & 0.025 $\pm$ 0.004 & 0.008 $\pm$ 0.003   \\\midrule
       kT$_{bb1}$  & 0.287 $\pm$ 0.001 &  0.179 $\pm$ 0.002 & 0.164 $\pm$ 0.002 & 0.164 $\pm$ 0.001 & 0.158 $\pm$ 0.005 & 0.159 $\pm$ 0.003 & 0.164 $\pm$ 0.004 & 0.159 $\pm$ 0.003 & 0.147 $\pm$ 0.003  \\\midrule 
       kT$_{bb2}$  & --- \par &  0.482 $\pm$ 0.005 & --- \par & --- \par & --- \par & --- \par & --- \par & --- \par & --- \par \\\midrule
       kT$_{mbb}$  & --- \par & --- \par & --- \par & 0.680 $\pm$ 0.007 & --- \par & 0.668 $\pm$ 0.008 & 0.673 $\pm$ 0.008 & 0.663 $\pm$ 0.008 & 0.654 $\pm$ 0.008 \\\midrule
       kT$_{dbb}$  & --- \par & --- \par & 1.2 $\pm$ 0.1 & --- \par & --- \par & --- \par & --- \par & --- \par & --- \par \\\midrule 
       kT$_{seed}$  & --- \par & --- \par & --- \par & --- \par & 0.158 (coupled) & --- \par & --- \par & --- \par & --- \par \\\midrule 
       kT$_{e}$  & --- \par & --- \par & --- \par & --- \par & 0.58 $\pm$ 0.02 & --- \par & --- \par & --- \par & --- \par  \\\midrule
       $\tau$  & --- \par & --- \par & --- \par & --- \par & 12 $\pm$ 2 & --- \par & --- \par & --- \par & --- \par  \\\midrule 
       kT$_{CIE}$  & --- \par & --- \par & --- \par & --- \par & --- \par & 1.11 $\pm$ 0.05 & 1.10 $\pm$ 0.05 &  --- \par & --- \par  \\\midrule 
       $N_{H}$  & 0.733 $\pm$ 0.001 & 2.23 $\pm$ 0.07 & 2.52 $\pm$ 0.07 & 2.53 $\pm$ 0.07 & 2.5 $\pm$ 0.1 & 2.44 $\pm$ 0.09 &  2.3 $\pm$ 0.1 & 2.40 $\pm$ 0.09 & 3.0 $\pm$ 0.2 \\\midrule
       ${N_H}_{Xabs}$  &  --- \par &  --- \par &  --- \par &  --- \par &  --- \par &  --- \par & 0.14 $\pm$ $^{0.26}_{0.10}$  &  --- \par &  --- \par   \\\midrule
       $ \rm Log \ \xi $ &  --- \par &  --- \par &  --- \par &  --- \par &  --- \par &  --- \par &  3.7 $\pm$ 0.1  &  --- \par &  --- \par  \\\midrule
       $E_{0}^{1}$  & --- \par & --- \par & --- \par & --- \par & --- \par & --- \par & --- \par & 0.98 $\pm$ 0.01 & 1.01 $\pm$ 0.02  \\\midrule
       $E_{0}^{2}$  & --- \par & --- \par & --- \par & --- \par & --- \par  & --- \par & --- \par & --- \par & 0.76 $\pm$ 0.01 \\\midrule
        FWHM  & --- \par & --- \par & --- \par & --- \par & --- \par & --- \par & --- \par &  0.22 $\pm$ 0.03 & 0.12 $\pm$ 0.03 \\\midrule
       $\rm Norm_{1}$  & --- \par & --- \par & --- \par & --- \par & --- \par & --- \par & --- \par & 1.6 $\pm$ 0.3 & 0.5 $\pm$ 0.2 \\\midrule
       $\rm Norm_{2}$  & --- \par & --- \par & --- \par & --- \par & --- \par & --- \par & --- \par & --- \par & -2.7 $\pm$ $^{0.7}_{1.0}$   \\\midrule
       ${\chi}^2$/d.o.f & 4537/287 & 463/285 & 437/285 &  428/285 & 438/284 & 360/283 & 342/281 & 346/282 & 325/279 \\\midrule 
       ${\chi}^2_{PN}$ & 2368 & 214 & 204 & 202 & 206 & 145 & 133 & 132 & 120  \\\midrule
       ${\chi}^2_{MOS1}$ & 1064 & 141 & 125 & 122 & 122 & 118 & 117 & 123  & 112  \\\midrule
       ${\chi}^2_{MOS2}$ & 1104 & 108 & 108 & 109 & 110 & 96 & 91 & 91 & 92 \\\midrule
    \bottomrule
    \label{table: XMM 0655050101 spectral fits}
    \end{tabular}}
    
    \begin{quotation}\footnotesize
    Parameter units: $E_{0}^{1}$ and $E_{0}^{2}$ (in keV unit) refer to the centroids of the first and second gaussian, respectively. 
    $\rm Norm_{1}$ and $\rm Norm_{2}$ (in $10^{46} \rm ph/s $ units) refer to the normalisations of the first and second gaussian, respectively. The temperatures kT (for each model) and FWHM are expressed in keV unit. The X-ray and bolometric luminosities $L_{X}$ and $L_{bol}$  (always intrinsic or unabsorbed) are calculated, respectively, between the 0.3 - 10 keV and 0.001 - 1000 keV bands, and are expressed in $10^{39}$ erg/s unit. The ionisation parameter $ \rm Log \ \xi $ is in  erg/s cm. The column density of the cold gas $N_{H}$ is in 10$^{21}$/cm$^{2}$ unit, while ${N_H}_{Xabs}$ is in 10$^{24}$/cm$^{2}$ unit. $\tau$ is the optical depth of the Comptonization component.
\end{quotation}
 \vspace{-0.3cm}
\end{table*}
\end{center}

\subsection{Spectral modelling of individual observations}
\label{Sec:spectra}

In Fig. \ref{fig: EPIC SPECTRA} we show the EPIC-PN spectra of the 10 XMM observations (the other two EPIC MOS 1,2 cameras are not shown here for clarity). We fitted the EPIC PN and MOS 1,2 from all observations using only continuum models because in some observations the residuals were too weak to constrain the parameters of the wind model. Besides the inclusion of the wind does not alter the continuum parameters even for the high quality spectra.

As for observation 06555050101, the spectral modelling was performed again simultaneously to the EPIC-PN, MOS1,2 spectra for each observation and testing all models mentioned before (blackbody, modified blackbody, powerlaw, and Comptonization). However, for clarity purposes and means of comparison with previous work, we focussed on the results obtained with a double thermal model. A warm ($kT_{bb}$ $\sim$ 0.2 keV) blackbody component reproduces the soft X-ray emission, presumably from the outer disc and the wind photosphere.
A hot ($kT_{mbb}$ $\sim$ 0.7 keV) blackbody component modified by coherent Compton scattering, accounts for the inner disc emission. The results are shown in detail in Table \ref{Results RHBM model} and the plots for the four observations with longest ($\gtrsim 90$ ks net) exposure and, therefore, the spectral fits are shown in Fig. \ref{fig: Spectral fit for 06555050101 observation using the RHBM model} (middle panel) and Fig. \ref{fig: Fit model for all observations using the RHBM model}.

This model provided a good description of the data except for the usual strong, narrow, residuals around 1 keV due to the well-known winds. Indeed, by adding a wind photoionisation absorption or collisionally ionised components - as previously found with the high-resolution RGS detectors - we could get rid of most spectral residuals (see observation 06555050101 as an example in Fig. \ref{fig: Spectral fit for 06555050101 observation using the RHBM model} bottom panel and table \ref{table: XMM 0655050101 spectral fits}). Our results are fully consistent with previous work and, in particular, seem to indicate that both thermal components become increasingly hotter (i.e. with a higher kT) at higher luminosities (\citealt{Pintore_2015,Pinto_2017}). For a comparison with results from the literature see Sect. \ref{Discussion}. 

The neutral column density was consistent among all high-quality, deep ($\gtrsim 100$ ks) exposures. The shortest ($\lesssim 30$ ks) exposures did not have sufficient signal-to-noise ratio to constrain both the spectral shape and the neutral ISM absorption. In particular, testing several models from single to multiple components model (RHB, RHBB, RHBM, etc.), we obtained values ranging from $(2-3) \times 10^{21} \rm cm^{-2}$. This was likely due to degeneracy produced by the lower statistics. Therefore, whilst fitting the spectra of these six short exposures we preferred to fix the $N_{H}$ to the average value obtained in the latter observations as previously done in e.g. \citet{Robba2021} for NGC 1313 ULX-2.

%		\begin{figure}[H]
%		\centering
%		\includegraphics[width=0.60\textwidth]{Pictures/comparison_spectra_new.PNG}
%        \vspace{-0.3cm}
%		\caption{{\small EPIC-PN spectra for all XMM observation of NGC 55 ULX-1. }}
%		\label{EPIC-PN spectra for all XMM observation of NGC 55 ULX-1. }
%		\end{figure}

		\begin{center}
	\begin{table*}%[h]
\caption{Results from the spectral modeling for the individual observations (RHBM model). }  
 \renewcommand{\arraystretch}{1.}
 \small\addtolength{\tabcolsep}{0pt}
 \vspace{0.1cm}
	\centering
	\scalebox{0.8}{%
	\begin{tabular}{ccccccccccc}
    \toprule
   % \multicolumn{5}{c}{{table model rhbm}} \\
   % \midrule
%   \multicolumn{6}{c}{{Best fit parameters of RHBM model}} \\
    % \midrule
    {{Parameter /}}  &
    {{0028740201}}  &
    {{0028740101}}  & {{0655050101}} &
    {{0824570101}} & {{0852610101}} & {{0852610201}} & {{0852610301}} & {{0852610401}} & {{0864810101}} & {{0883960101}}\\
    {{component}} & {{Obs}}  &  {{Obs}}   & {{Obs}}   & {{Obs}}   & {{Obs}}   & {{Obs}}  & {{Obs}} & {{Obs}} & {{Obs}} & {{Obs}}\\ 
    
    \midrule
       $L_{X\,bb}$      & 1.50 $\pm$ 0.06   & 1.12 $\pm$ 0.08   &  0.93 $\pm$ 0.02   &  0.71 $\pm$ 0.02   & 1.5 $\pm$ 0.2     & 1.6 $\pm$ 0.2     & 0.7 $\pm$ 0.1     & 1.1 $\pm$ 0.2     &  0.91 $\pm$ 0.02  & 1.15 $\pm$ 0.03 \\\midrule
       $L_{X\,mbb}$     & 1.37 $\pm$ 0.06   & 1.48 $\pm$ 0.08   &   0.57 $\pm$ 0.02  & 0.38 $\pm$ 0.01    & 1.2 $\pm$ 0.2     & 1.2 $\pm$ 0.2     & 0.46 $\pm$ 0.03   & 0.6 $\pm$ 0.2     & 0.58 $\pm$ 0.02   & 0.67 $\pm$ 0.02 \\\midrule
       kT$_{bb}$        & 0.172 $\pm$ 0.002 & 0.174 $\pm$ 0.003 & 0.164  $\pm$ 0.001 & 0.162 $\pm$ 0.001  & 0.168 $\pm$ 0.004 & 0.162 $\pm$ 0.004 & 0.153 $\pm$ 0.005 & 0.163 $\pm$ 0.006 & 0.167 $\pm$ 0.001 & 0.166 $\pm$ 0.001\\\midrule
       kT$_{mbb}$       & 0.82 $\pm$ 0.01   & 0.90 $\pm$ 0.02   & 0.680 $\pm$ 0.007  & 0.755 $\pm$ 0.008  & 0.71 $\pm$ 0.03   & 0.69 $\pm$ 0.03   & 0.67 $\pm$ 0.04   & 0.68 $\pm$ 0.05   & 0.713 $\pm$ 0.007 & 0.680 $\pm$ 0.006 \\\midrule 
       $N_{H}$          & 2.5               & 2.5               & 2.53 $\pm$ 0.07    & 2.57 $\pm$ 0.09    & 2.5               & 2.5               & 2.5               & 2.5               & 2.52 $\pm$ 0.07   &  2.5 \\\midrule
       ${\chi}^2$/d.o.f & 334/275           & 294/240           & 428/285            & 413/279            & 192/153           & 139/137           & 135/97            & 87/87             & 463/285           & 378/294 \\\midrule
       $L_{bol}$        & 3.19              & 2.85              & 1.71               & 1.26               & 3.15              & 3.17              & 1.32              & 2.01              & 1.69              & 2.08 \\\midrule
    \bottomrule
    \label{Results RHBM model}
    \end{tabular}}
    \begin{quotation}\footnotesize
       Units are the same as in the Table \ref{table: XMM 0655050101 spectral fits}.
\end{quotation}
        \vspace{-0.3cm}
\end{table*}
\end{center}

\begin{figure}%[H]
		\centering
		\renewcommand{\arraystretch}{1.}

%       \includegraphics[width=0.45\textwidth]{Images/PLT_0028740101_all_final.pdf}
%       \includegraphics[width=0.45\textwidt%h]{Images/PLT_00287%40201_all_final.pdf}

%        \vspace{-0.5cm}
%		\label{fig: Fit model for all observations using the RHBM model (1) }
%		\end{figure}
%		
%		\begin{figure}[H]
%		\centering
        \includegraphics[width=0.45\textwidth]{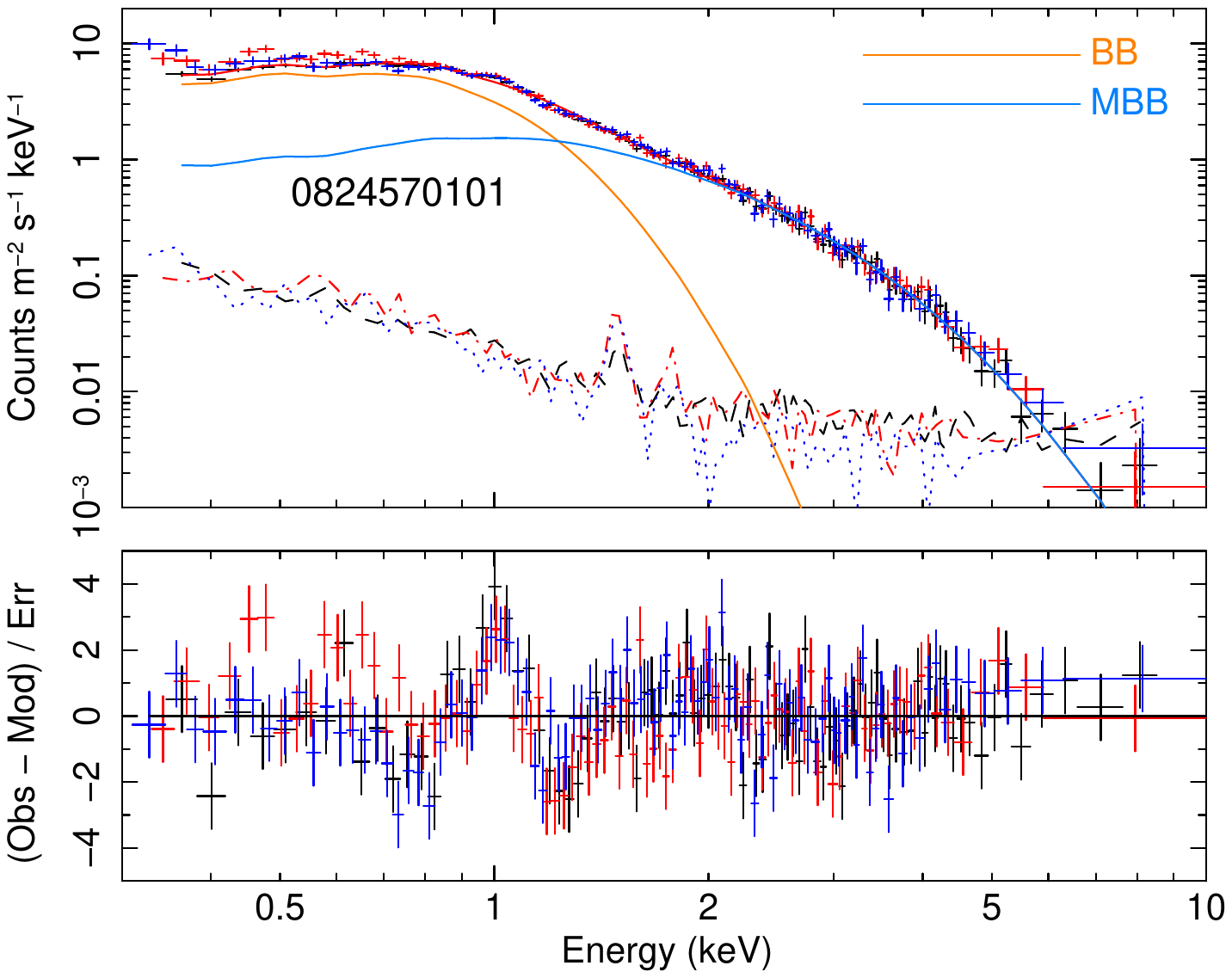}
        \includegraphics[width=0.45\textwidth]{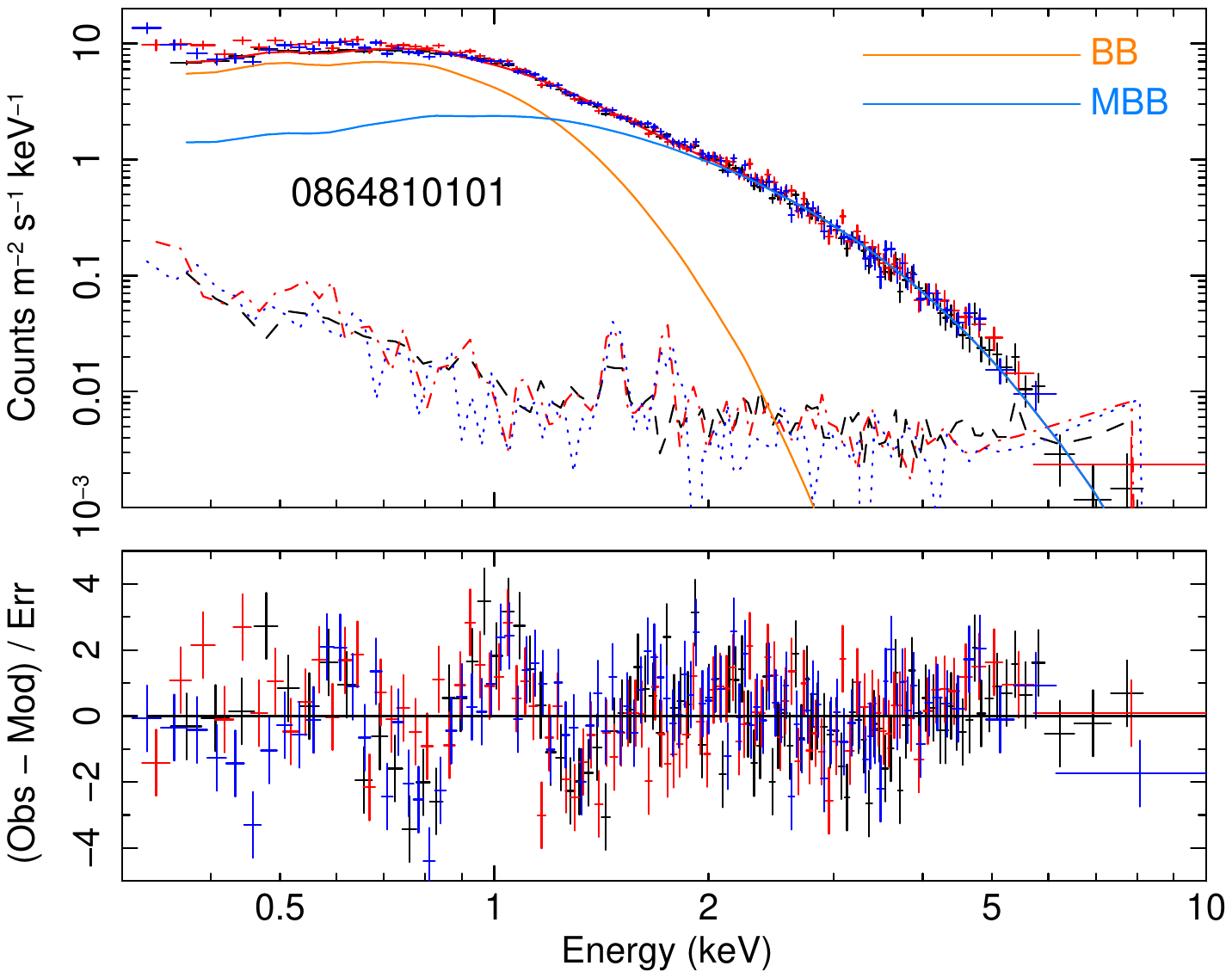}
        \includegraphics[width=0.45\textwidth]{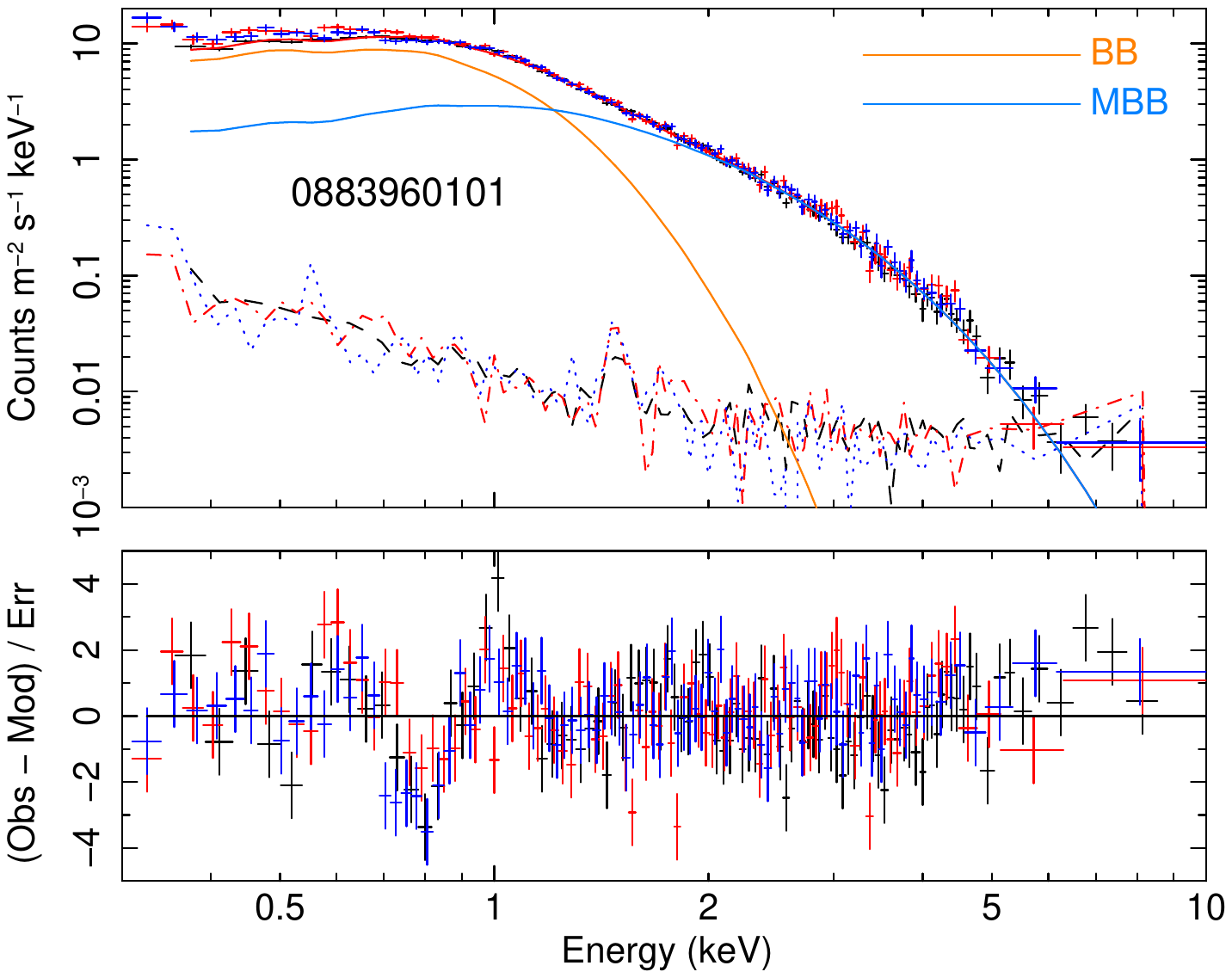}
        
        \vspace{-0.3cm}
		\caption{{\small Spectral fits for the other three long ($\gtrsim 90$ ks net) observations of NGC 55 ULX-1 using the RHBM model (blackbody + modified blackbody). }}
		\label{fig: Fit model for all observations using the RHBM model}
        \vspace{-0.3cm}
		\end{figure}

\subsection{Spectral modelling of different flux levels}

In order to confirm and corroborate any trends between spectral hardness and source flux we fit the eight flux-resolved spectra shown in Sect. \ref{Observations and Spectral modelling} with the RHBM model as done before for the individual observations. The results of the spectral fits, fixing the column density $N_{H}$ to the average value obtained in the previous results ($ 2.5 \times 10^{21} \rm cm^{-2}$), are shown in table \ref{Results RHBM model-FRS}. At higher luminosities, the temperature of the hot modified blackbody component increased from 0.70 to 0.86 keV while the one for the cooler component increased by just 0.01 keV. An exception was the lowest flux level 1, which was largely dominated by Obs. ID: 0824570101, where there seemed to be an increase in the \textit{mbb} temperature. An interesting result is the consistency between the parameters of the cool \textit{bb} component between the highest level 8 and the bright, dipping, level 7. At odds, there was instead a clear drop in the luminosity and temperature of the hot \textit{mbb} component during level 7. The results from these spectral fits are discussed later on in Sect. \ref{Discussion}.

In Sect. \ref{sec:wind_variability} we show how gaussian lines can describe the spectral residuals in these high-quality flux-resolved spectra, discuss their variability and how they may provide insights on the overall disc evolution and structure.

		\begin{center}
	\begin{table*}%[h]
\caption{Results from the modeling with the RHBM model of the eight flux-resolved spectra.}  
 \renewcommand{\arraystretch}{1.}
 \small\addtolength{\tabcolsep}{0pt}
 \vspace{0.1cm}
	\centering
	\scalebox{0.9}{%
	\begin{tabular}{ccccccccc}
    \toprule
   % \multicolumn{5}{c}{{table model rhbm}} \\
   % \midrule
%   \multicolumn{6}{c}{{Best fit parameters of RHBM model}} \\
    % \midrule
    {{Parameter /}}  &
    {{Level}}  &
    {{Level}}  & {{Level}} &
    {{Level}} & {{Level}} & {{Level}} & {{Level}} & {{Level}} \\
    {{component}} & {{1}}  &  {{2}}   & {{3}}   & {{4}}   & {{5}}   & {{6}}  & {{7}} & {{8}} \\ 
    
    \midrule
     $L_{X\,bb}$  & 0.70 $\pm$ 0.02 & 0.81 $\pm$ 0.03   &  0.90 $\pm$ 0.03 &  1.00 $\pm$ 0.03 &  1.11 $\pm$ 0.03   & 1.20 $\pm$ 0.04 & 1.39 $\pm$ 0.05 & 1.40 $\pm$ 0.07 \\\midrule
    $L_{X\,mbb}$  & 0.38 $\pm$ 0.02 & 0.50 $\pm$ 0.02   &   0.55 $\pm$ 0.03 & 0.60 $\pm$ 0.03 & 0.65 $\pm$ 0.03 & 0.70 $\pm$ 0.04 & 0.89 $\pm$ 0.05  & 1.65 $\pm$ 0.07 \\\midrule
       kT$_{bb}$ & 0.162 $\pm$ 0.001 & 0.162 $\pm$ 0.001  & 0.165  $\pm$ 0.001 & 0.166 $\pm$ 0.001 & 0.166 $\pm$ 0.001 & 0.167 $\pm$ 0.001 & 0.170 $\pm$ 0.001 & 0.175 $\pm$ 0.002 \\\midrule
       kT$_{mbb}$ & 0.75 $\pm$ 0.01 & 0.704 $\pm$ 0.008  & 0.705 $\pm$ 0.009  & 0.694 $\pm$ 0.009 & 0.692 $\pm$ 0.009 & 0.669 $\pm$ 0.009 & 0.74 $\pm$ 0.01 & 0.89 $\pm$ 0.01 \\\midrule 
       $N_{H}$ & 2.5 & 2.5 & 2.5 & 2.5 & 2.5 & 2.5 & 2.5 & 2.5 \\\midrule
       ${\chi}^2$/d.o.f & 452/282 & 431/276  & 390/279 & 350/274 & 366/271 & 295/262 & 324/270 & 390/295 \\\midrule
       $L_{bol}$ & 1.24 & 1.49 & 1.65 & 1.82 & 2.00 & 2.17 & 2.58 & 3.36 \\\midrule
    \bottomrule
    \label{Results RHBM model-FRS}
    \end{tabular}
    }
    \begin{quotation}\footnotesize
     Units are the same as in the Table \ref{table: XMM 0655050101 spectral fits}.
\end{quotation}
        \vspace{-0.3cm}
\end{table*}
\end{center}

\subsection{Variability of the wind features}
\label{sec:wind_variability}

%....to be continued...

%\textcolor{red}{CP: we should compare the strength of the lines with NGC 247. Here normalisation is 150 -- 200 for the emission line at 1 keV and -100 -- -200 for 0.8 keV and -30 -- -70 for the 1.25 keV feature.}
In order to study the strength of the spectral features and their variability in NGC 55 ULX-1, we performed a quick fit of the flux resolved spectra modifying our baseline \textit{bb} + \textit{mbb} model by adding three gaussian lines (RHBMGGG model). Following \citet{Pinto_2021}, we fixed the line width to 1 eV for all three gaussian lines since EPIC spectra lack the necessary spectral resolution around 1 keV. The energy centroids were free to vary however they agreed within the error bars with 0.8 and 1.2 keV for the absorption lines and 1.0 keV for the emission line.

\begin{figure}%[ht!]
		\centering
		\includegraphics[width=0.45\textwidth]{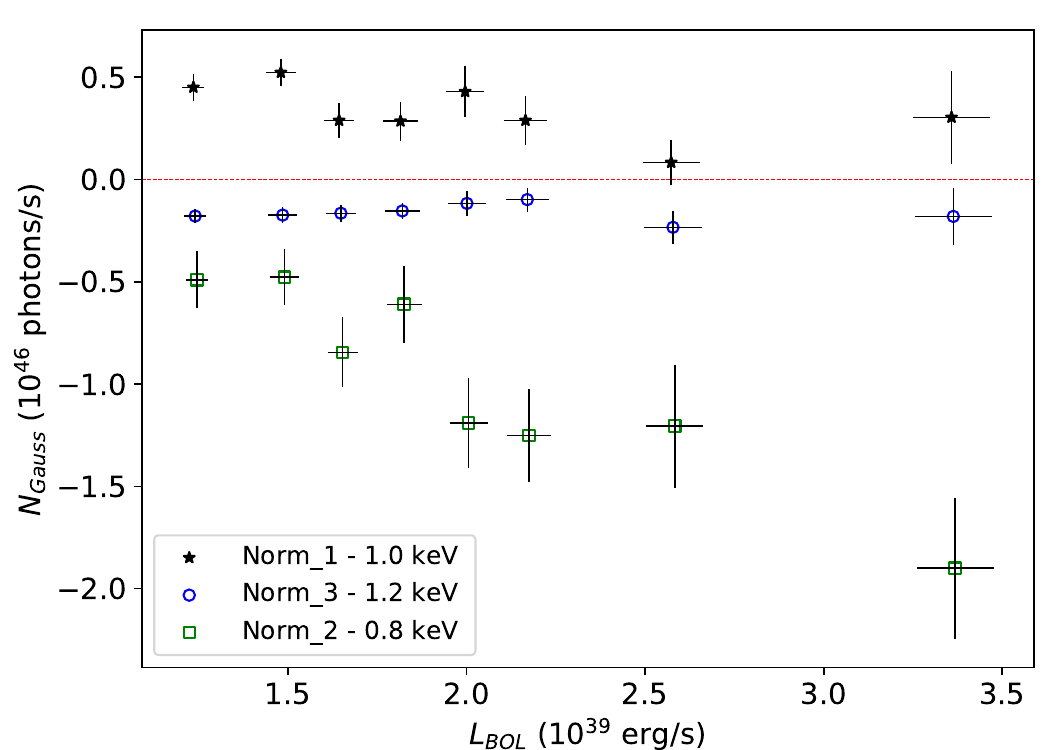}
        \vspace{-0.1cm}
		\caption{{Normalisation of the gaussian lines vs Bolometric luminosity for the 3-gaussian model fits of the flux resolved spectra. The negative normalisations refer to absorption lines.}}
		\label{fig: Normalization_vs_Bolometric Luminosity plot}
		\end{figure}

The normalisations of the gaussian lines for the eight spectra are shown in Fig. \ref{fig: Normalization_vs_Bolometric Luminosity plot}. They do not show strong trends with the flux, possibly due to fairly large uncertainties, with the exception of the lowest-energy line at 0.8 keV, which seems to get stronger at high fluxes in agreement with NGC 1313 ULX-1 (\citealt{Pinto_2020b}) and NGC 247 ULX-1 (\citealt{Pinto_2021}). These trends were also attributed to an increasing $\dot M$.

%We can compare the strength of the lines of this source with those measured in NGC 247 ULX-1 at the same energies \citep{Pinto_2021}. For NGC247 ULX-1, the lines are characterized by a normalisation of $1.5-2.0\times10^{46}$ photons/s for the emission lines at 1 keV, -1 to -2 $\times10^{46}$ photons/s for 0.8 keV feature and\textcolor{red}{ -0.3 to -0.7 $\times10^{46}$ photons/s for the 1.2 keV feature. As can be seen from Fig. \ref{fig: Normalization_vs_Bolometric Luminosity plot}, such values are 2-3 times larger than those found in NGC 55 ULX-1 (even for the emission lines). This would suggest an intrinsically more powerful wind in NGC 247 ULX-1 as a consequence of a higher accretion rate in agreement with the continuum properties discussed in Sec. \ref{sec:comparison_with_other_ulxs}.}

\section{Discussion}
\label{Discussion}

%%%\textcolor{blue}{Provide a quick summary of the results (and maybe mention some of the aims) then discuss the results and compare to other ULXs and previous work by others.} \\
%%%\textcolor{red}{QUICK SUMMARY OF THE RESULTS}

The aim of this work is to {investigate} the processes that trigger the spectral transitions in ULXs, particularly between the soft and intermediate hardness regimes. The main question is whether they are driven by stochastic variability in the wind (e.g. \citealt{Kobayashi2018}) or variations in the accretion rate, which in turn produce variability in the wind and in the obscuration of the innermost, hottest, regions (e.g. \citealt{Urquhart2016}). Geometrical effects may also play a relevant role in systems where the accretion disc precesses (e.g. \citealt{Middleton_2015a}).

According to {radiation-magnetohydrodynamic (r-MHD)} simulations of super-Eddington accretion discs (e.g. \citealt{Takeuchi2013}), the winds are expected to become optically thick enough to block and reprocess a fraction of the disc X-ray photons, making the source appear as a soft thermal emitter or ultraluminous (super)soft X-ray source (e.g. \citealt{Guo2019}) {at moderately high inclination angles}. A comprehensive study would require to measure the properties of the winds 
%(e.g. column density, temperature and velocity) 
and compare them to the characteristics of the spectral continuum 
%(luminosity, number of significant components, temperatures) at different flux regimes 
(similarly to the \citealt{Pinto_2020b} for the intermediate-to-hard source NGC 1313 ULX-1). In order to do this, we used the XMM-\textit{Newton} observations of one of the most variable and brightest (in flux) nearby source NGC 55 ULX-1. The X-ray spectra of this ULX fit in just between the softest and intermediate ULX spectra (see Fig. \ref{fig: Comparison X-ray spectra of the brightest ULX known}). Indeed, the temperatures of the \textit{bb} fits are among the lowest in ULXs (e.g. \citealt{Sutton_2013}). Moreover, we benefited from three new deep observations that enabled us to achieve more statistics in the low-flux regime and 4 new short observations which fill the gaps between the high- and low-flux regimes. In this work we focused on the evolution of the spectral continuum throughout the different epochs. A follow-up work will focus on the study of the high-resolution X-ray spectra which are however available only for the low-intermediate flux on-axis observations and the evolution of the wind in detail.

%%% In addition to XMM, we have also used data from the Swift satellite which allows brief but frequent visits of any source in the sky. The analysis of the source was therefore performed using the Swift and XMM-\textit{Newton} satellites for timing (long and short timescale monitoring, years and hours, respectively) and X-ray spectroscopy. The long monitoring with SWIFT/XRT shows evidence for a strong variability by up to a factor 4 in flux over timescales of a few days. 

\subsection{X-ray broadband properties}

The XMM-\textit{Newton}/EPIC lightcurves showed evidence for strong variability by up to a factor 4 in flux over timescales of a few days (see Fig. \ref{fig: lightcurves}, top panel) along with dips lasting a few 100s to a few ks, which have been interpreted as due to wind clumps that temporarily obscured the central object (see, e.g., \citealt{Stobbart_2004,Pintore_2015}). In fact, during the dips the spectral hardness decreased and reached levels comparable to the low-flux observations although the dip flux was 2-3 times higher than that of the low-flux observations (Fig. \ref{fig: lightcurves}, bottom panel). The dips appear mainly when the source is bright indicating that the accretion rate increases thereby launching optically-thick wind clouds into the LOS. This suggests the presence of at least two different ongoing variability processes (see also below). \textit{Swift}/XRT long-term lightcurves taken between 2013 and 2021 confirm the general source behavior with flux variations by a factor of up to 6 with the spectrum of the source appearing harder when brighter \citep{Jithesh_2021}.

We extracted spectra for individual observations to probe variations on timescales of a few days to years as well as flux-resolved spectra to probe the nature of on {different timescales} variability (see Fig. \ref{fig: EPIC SPECTRA} and \ref{fig: EPIC SPECTRA FRS}). 
Flux level 1 mainly sampled the lowest flux observation (Obs. ID:082457), while flux levels 7 and 8 described the dip and no-dip time intervals in the bright epochs (Obs. ID: 002874-0201, from 2001, see Fig. \ref{fig: lightcurves}). The other levels (2-6) traced the intermediate-flux epochs (Obs. ID: 065505,086481). All spectra peaked around 1 keV with the peak slightly shifting towards higher energies at higher fluxes. Interestingly, the lowest-flux spectrum appeared harder than the intermediate-flux ones {in contradiction with the trend sees at higher fluxes.}

%%%We first focused on the archival Obs. ID: 0655050101 in order to decompose the spectrum into the main continuum components and to choose the baseline continuum model for all the other observations. Wind models were also tested to model the narrow residuals and to understand the effects of the wind inclusion onto the continuum parameters.
Our analysis confirmed that the spectrum requires at least two components with a blackbody-like shape. A double blackbody model could indeed reproduce the overall spectral shape with the hotter component being broader and likely modified by Compton scattering in the wind (see Fig. \ref{fig: Spectral fit for 06555050101 observation using the RHBM model} and Table \ref{table: XMM 0655050101 spectral fits}). This agrees with the general picture and the typical properties of soft ULXs (see, e.g., \citealt{Sutton_2013}). The inclusion of wind features in the model in the form of gaussian lines or physical components %%%(e.g. collisional ionisation emission or photoionised absorption) 
produced significant improvements to the spectral fits in terms of $\Delta \chi^2$ but did not affect the continuum parameters
albeit slightly increasing the uncertainties (see Table \ref{table: XMM 0655050101 spectral fits}). 
%%%This means that we could just use continuum models to fit the observations, which certainly helped since some of them are very short and did not provide sufficient statistics to constrain the wind properties (a similar approach was used by \citealt{Walton_2020}).

The spectral fits of both the individual observations and the eight flux levels showed that both the soft and hard components became hotter (i.e. with a higher kT) at higher luminosities (see Table \ref{Results RHBM model} and \ref{Results RHBM model-FRS}), which indicates that either the accretion rate is increasing or that we are progressively having a clearer view of the inner accretion flow through a less dense wind photosphere. However, the results on the wind features (e.g. the more intense 0.8 keV line at high fluxes) would favor an increase in the ${\dot M}$ (see Sect. \ref{sec:wind_variability}).
This agrees with a scenario in which variations in the local accretion rate are driving the transition from soft/fainter to hard/brighter spectra. The brightest ones are associated to epochs in which a surplus of matter may launch optically-thick clouds that obscure the innermost region thereby producing the flux dips.

\subsection{Comparison with disc models}

The nature of the compact object powering NGC 55 ULX-1 is not yet known and therefore the accretion in Eddington units is uncertain. Although the best-fit values of temperatures might not exactly correspond to gas temperatures due to the unknown structure of the emitting regions {(because we have modelled a thick disk with a thin disk template)}, it is still very useful to compare them with the bolometric luminosities of the blackbody components. This is indeed a commonly adopted procedure to understand the behaviour of the thermal components and, in particular, to place some constraints on the disc structure and accretion regime, possibly providing some information on the nature of the compact object, especially for not too high accretion rates or luminosities ($L_{\rm BOL}\sim10^{39}$ erg/s, e.g., \citealt{Urquhart2016,Earnshaw2017,Walton_2020,Gurpide_2021a,Robba2021,D'Ai_2021}). 
%%% The X-ray luminosity of NGC 55 ULX-1 is just above $10^{39}$ erg/s, which means that - unless the innermost regions are highly obscured - it could bear a rather thin or slim disc.

It is useful to compare the Luminosity--Temperature ($L-T$) trends measured for NGC 55 ULX-1 with those expected from theoretical models such as the thin disc in a sub-Eddington regime (L $\propto$ $T^{4}$) with a constant emitting area (SS73, \citealt{SS1973}) and the advection-dominated disc model with L $\propto$ $T^{2}$ (\citealt{10.1093/pasj/53.5.915}).
In Fig. \ref{fig: L-T plot} (top panel) we show the trends between the temperature and the bolometric luminosity for both the cool (blue points) and warm (orange points) blackbody components from the spectral fits of the ten individual XMM observations.
Overlaid are the best-fitting least-squares regression line (solid blue, $L \propto T^{\alpha}$), the L $\propto$ $T^{4}$ (dashed red) and the L $\propto$ $T^{2}$ (dotted black) trends. In Fig. \ref{fig: L-T plot} (bottom panel) we show the same results obtained with the eight spectra extracted in the count rate ranges of the XMM lightcurve (see Fig. \ref{fig: lightcurves}).
The bolometric (or total) luminosities of the cool blackbody and warm modified blackbody components are computed between 0.001-1000 keV by extrapolation, although {most of the flux is emitted in the 0.1-10 keV range}. Both plots show that the measured L--T relationship is in broad agreement with the thin SS73 disc.
{From the regression lines we obtain a power index, with the time resolved spectroscopy $\alpha_{cool}= 4.2 \pm 1.9 $ and $\alpha_{hot}= 6.1 \pm 2.3 $, for the cool and hot components, respectively. Instead, with the flux resolved spectroscopy, the power indices are $\alpha_{cool}= 5.4 \pm 1.7 $  and $\alpha_{hot}= 5.4 \pm 1.8 $. The possible correlation of the points in the L--T plot, for the cool and hot components, can be established with the Pearsons and Sperman correlation coefficients. These are reported in the Table \ref{pearsons and spearman correlation coefficients} by using \textit{scipy.stats.pearsonsr} and \textit{scipy.stats.spearmanr} routine in {{\scriptsize PYTHON}. They do not show always strong correlations.}}
\begin{center}
	\begin{table}
	\caption{Pearsons and Spearman coefficients for the L-T trends, for both cool and hot components, by using the best fit values of the RHBM model.}  
	 \renewcommand{\arraystretch}{1.}
 \small\addtolength{\tabcolsep}{0pt}
 \vspace{0.1cm}
	\centering
	\scalebox{0.85}{%
  \begin{tabular}{c c c c c}
    \hline
    \multirow{2}{*}{Correlation coefficient} &
      \multicolumn{2}{c}{TRS}  & 
      \multicolumn{2}{c}{FRS} \\ \cline{2-5}
    &  ${(L-T)}_{cool}$ & ${(L-T)}_{hot}$ & ${(L-T)}_{cool}$ & ${(L-T)}_{hot}$  \\
    \hline
    Pearsons & 0.48 & 0.65 & 0.90 & 0.83 \\
    \hline
    Spearman & 0.39 & 0.55 & 0.93 & 0.05  \\
    \hline
  \end{tabular}}   \label{pearsons and spearman correlation coefficients}
  \begin{quotation}\footnotesize
TRS and FRS acronyms stand for time-resolved spectroscopy and flux-resolved spectroscopy, respectively.
\end{quotation}
        \vspace{-0.3cm}
\end{table}
\end{center}
Locally there are small deviations in Fig. \ref{fig: L-T plot} from the tight correlations. 
%%%This is probably due to the fact that the intrinsic luminosity of the source is not very high and, therefore, the accretion rate is just mildly super-Eddington. This result would suggest a black hole rather than a neutron star accretor.
In particular, at high luminosities the temperature of the cool component is lower than the predictions from the L $\propto$ $T^{4}$ model or the regression line, which suggests that the disc is expanding and, perhaps, that the wind starts to contribute to the emission {or the radius of the thermal component is not constant}. The lowest-flux observation also shows a notable deviation, especially for the warm disc component, with a temperature higher than as predicted from the $L-T$ trends. This could be a hint of low/hard - high/soft behavior seen in Galactic X-ray binaries and might indicate a spectral transition below $10^{39}$ erg/s, {although the spectra of NGC 55 ULX-1 never get as hard as the Galactic X-Ray Binaries (XRBs) likely due to its persistently high accretion rate} (\citealt{Koljonen_2010}).

 Fig.\,\ref{fig: L-T plot} shows that there are deviations for the luminosity and temperatures from the best-fit regression line, approximately for {total} luminosity greater than $2 \times 10^{39} \rm erg/s$. This fact can be used to give a rough estimate of the mass of the compact object. For instance, if we assume that the deviation is due to the total luminosity reaching the Eddington limit ($L_{\rm tot} \sim L_{\rm Edd}$) and that the apparent luminosity is comparable to the intrinsic luminosity, using the definition $L_{\rm Edd} = 1.4 \times 10^{38} \frac{M}{M\textsubscript{\(\odot\)} }$ erg/s, we estimate that the mass of the compact object is about 14 $M\textsubscript{\(\odot\)}$. Instead, if we assume that the deviation is due to the disc becoming supercritical ($L_{\rm tot}\sim L_{\rm critical} = 9/4 \  L_{\rm Edd}$, \citealt{Poutanen_2007}), then we obtain a value for the mass of the compact object of about 6 $M\textsubscript{\(\odot\)}$. Of course, should the intrinsic luminosity be greater than the apparent luminosity, which is possible given the weakness of the hard component, then the mass of the compact object would be larger. Bright ULXs with hard spectra typically have $L_{\rm Bol} = (0.5-1) \times 10^{40}$ erg/s, which would result in 10--30 $M\textsubscript{\(\odot\)}$ for NGC 55 ULX-1. Therefore, the  mass range obtained would suggest that the compact object is a black hole. This agrees with the results obtained by \citet{Fiacconi_2017} using wind arguments. {Interestingly, some deviations are also seen in Galactic XRBs above $0.3 \ L_{Edd}$ (e.g. \citealt{Steiner_2009}). If the deviations that we see refer to such threshold, then a larger mass of the compact object, i.e. a heavier BH, is forecast.}
 
 The radius of the cool component has a value of the order of $3000 \rm \ km$; instead the radius of the hot component is of the order of $100 \rm \ km$, ({estimated with the relations between luminosity and temperature and blackbody definition}). These correspond to, respectively, 200 $R_{G}$ and 10 $R_{G}$, assuming a black hole of 10 $M\textsubscript{\(\odot\)}$. The radii of both components from the first to the eighth flux level do not vary significantly possibly due to the much larger uncertainties with respect to the temperature. Such results show a slight tension with \citet{Jithesh_2021}
in which a tentative anti-correlation between the radius and the temperature of the cool component was found albeit at large uncertainties. This is probably due to the fact that we fixed the column density $\rm N_{H} $ since we do not expect a strong variation of the neutral gas within a few hours, {which are the timescale of the dips and the separation between consecutive observations}.

By considering that the spherisation radius is $R_{\rm sph}= 27/4 \ \dot{m} \ R_{G}$ and assuming that $\dot{m}$ $\sim$ $\dot{m}_{Edd}$, the radius of the hot blackbody component is comparable to the spherisation radius ($R_{\rm sph}\sim 7 R_{G}$) (\citealt{Poutanen_2007}), instead the radius of the cool component is significantly larger than $R_{\rm sph}$ identifying the outer disc. However, these results are correct if the intrinsic luminosity is comparable with the observed one or if there are not large losses due to the occultation of the inner portion of the disc by the wind. \citet{Jithesh_2021} also used \textit{NuSTAR} data to better constrain the hard band; {we do no think the use of these data might alter our conclusions, since there is only marginal flux in the NuSTAR band which is not covered by the EPIC-PN}. Besides, there are only a few simultaneous  XMM/\textit{NuSTAR} observations.

\subsection{Comparison with other ULXs}
\label{sec:comparison_with_other_ulxs}

In the brighter ($L_X$ up to $10^{40}$ erg/s) and pulsating source NGC 1313 ULX-2 there is a clear anti-correlation between the bolometric luminosity and the temperature of the cool blackbody-like component (\citealt{Robba2021}). In that source the trend is in agreement with the L $\propto$ $T^{-4}$ relationship predicted by X-ray emission from a wind photosphere rather than a disc (e.g. \citealt{Qiu_2021}, \citealt{Kajava2009}, \citealt{King_2009}), and is likely due to the higher observed luminosity and possibly higher accretion rate for the pulsating neutron star. In the even brighter source in same galaxy, NGC 1313 ULX-1, \citet{Walton_2020} found a deviation in the L--T relationship for the warm disc-like component which has been attributed to an intervening wind or a higher disc scale height. The cool component seems fairly constant {in luminosity} in NGC 1313 ULX-1. The debate {on the nature of the different L-T trends} is still open.

\begin{figure}%[H]
		\centering
		\includegraphics[width=0.475\textwidth]{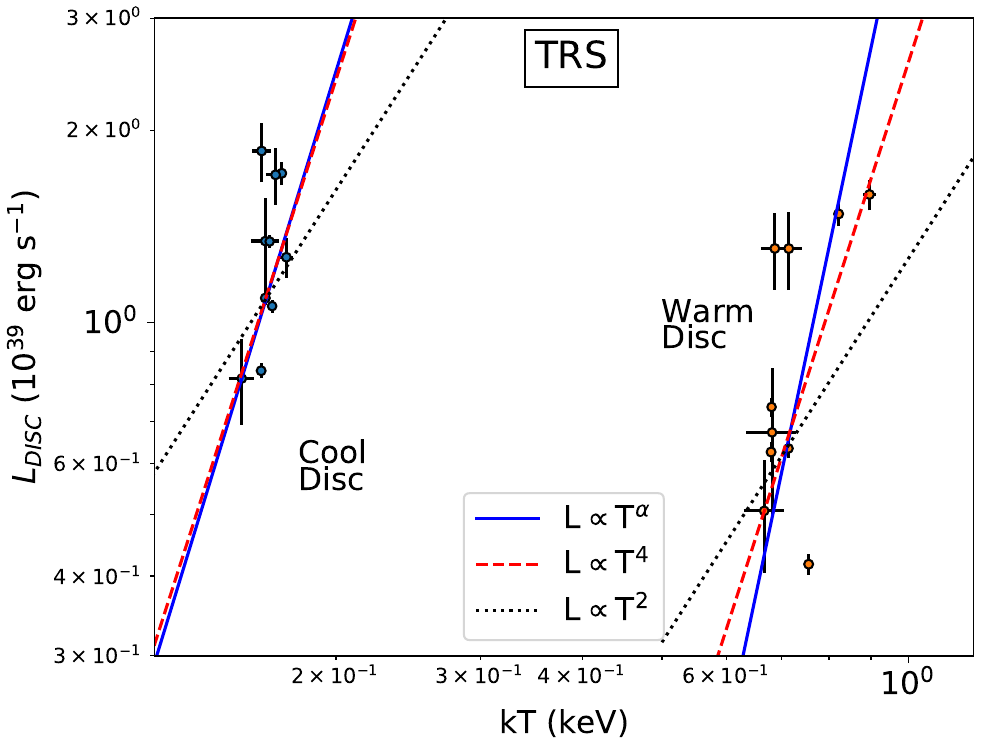}
		\includegraphics[width=0.475\textwidth]{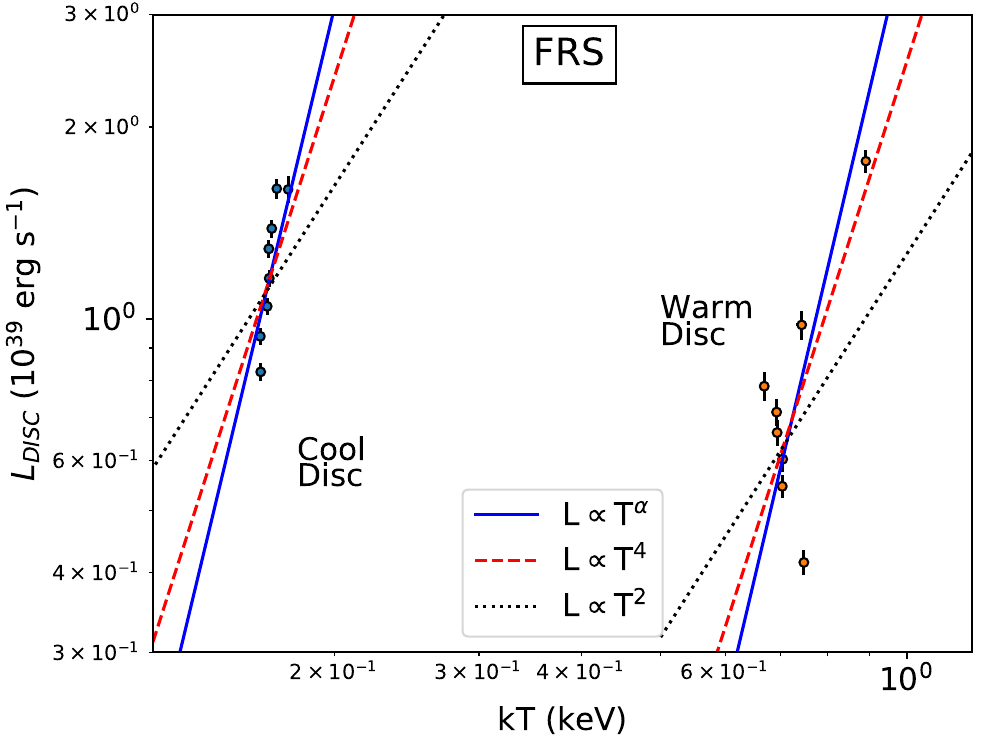}
        \vspace{-0.1cm}
		\caption{{\small Bolometric luminosity estimated in the 0.001-1000 keV energy band versus temperature for the cool blackbody (in blue) and hot modified blackbody (in orange) components. Top panel: Luminosity-temperature plot for the time resolved spectra (TRS). Bottom panel: Luminosity-temperature plot for the flux resolved spectra (FRS). The blue solid line, the black-dotted and the red dashed lines represent the regression line and the two theoretical models of the slim disc and Shakura-Sunyaev models, respectively. }}
		\label{fig: L-T plot}
		\end{figure}

The lightcurve of NGC 55 ULX-1 shows a dipping behavior during the high-flux epochs which is very similar to that one shown by supersoft ULXs such as NGC 247 ULX-1 (\citealt{Feng_2016}). There are much more data available for NGC 247 ULX-1 thanks to a recent deep XMM-Newton campaign (PI: Pinto). The results from the broadband analysis were shown by \citet{D'Ai_2021} who reported on a complex hardness-intensity diagram characterised by two main branches; here at high luminosities the source enters the dipping branch in which the spectrum becomes progressively softer. It is possible that the deviation of the temperature of the cool component shown by NGC 55 ULX-1 at high luminosities (see above and Fig. \ref{fig: L-T plot}) is the first stage of the dipping behavior seen in NGC 247 ULX-1 but with the wind not optically-thick enough to completely obscure the emission above 1 keV. This could be due to either a lower accretion rate or inclination angle in NGC 55 ULX-1. Given the lower luminosity and harder spectrum of NGC 55 ULX-1 as compared with the NGC 247 ULX-1 (see, e.g., \citealt{Pinto_2021}), it is reasonable to speculate that the former is at a lower accretion rate with a thinner wind in the line of sight.
Indeed, the dominant emission line {blend} at 1 keV from the wind is much stronger in NGC 247 ULX-1. More precisely, when modelled it with a \textit{cie} component we obtained $L_{\rm X \, [0.3-10 \, keV]} \sim 0.7 \times 10^{38}$ erg/s in NGC 55 ULX-1 (see Table \ref{table: XMM 0655050101 spectral fits}), while for NGC 247 ULX-1 it was $\sim 1.4 \times 10^{38}$ erg/s \citep{Pinto_2021}, i.e. twice as strong. 
A more powerful wind in NGC 247 ULX-1 is likely a consequence of a higher accretion rate.

\section{Conclusions}
\label{Conclusions}

In this work we have performed a spectral analysis to understand the structure and evolution of the accretion disc in the ultraluminous X-ray source NGC 55 ULX-1. The archival data was enriched with three deep XMM-\textit{Newton} observations that we obtained in 2018, 2020 and 2021. This enabled to follow up the {continuum changes and their relation} with the source luminosity. As for most ULXs, it is necessary to use at least two spectral components to reproduce the spectral shape. The spectrum can be well fit with a two-blackbody model composed of a cool blackbody component, that describes the softer X-ray emission coming from the outer and cool part of the accretion disc and the wind photosphere, and a hotter blackbody modified by Compton scattering, that instead accounts for the emission from the hot inner disc.  Both components become hotter at higher luminosities, indicating either a better and clearer view of the inner disc, due to e.g. a reduction of the wind photosphere density, or to an increase in the accretion rate. The variability of the wind features would favour the latter case. The trends between the bolometric luminosity and temperature of each component broadly agree with the $L \propto T^{4}$ relationship expected from constant area such as a thin disc. This suggests that the intrinsic luminosity of the source is not extremely high and likely close to the Eddington limit of a 10 $M\textsubscript{\(\odot\)}$ black hole. At high luminosities the cool component is cooler than the predictions from the thin-disc model and the best-fit regression line. This would imply an expansion of the disc and a contribution to the emission from the wind. If the deviation occurs between the Eddington limit and the supercritical accretion rate, a black hole within 6-14 $M\textsubscript{\(\odot\)}$ is foreseen.

\section{Data availability}

All of the data, with the exception of the Obs. ID 0883960101, and software used in this work are publicly available from ESA's XMM-\textit{Newton} Science Archive (XSA\footnote{https://www.cosmos.esa.int/web/XMM-\textit{Newton}/xsa}) and NASA's HEASARC archive\footnote{https://heasarc.gsfc.nasa.gov/}. 
%%%Our codes are publicly available and can be found on the %%%GitHub\footnote{https://github.com/ciropinto1982}.

\section*{Acknowledgements}

This work is based on observations obtained with XMM-\textit{Newton}, an ESA science mission funded by ESA Member States and USA (NASA). This work has been partially supported by the ASI-INAF program I/004/11/4 from the agreement ASI-INAF n.2017-14-H.0 and from the INAF mainstream grant. We acknowledge the XMM-\textit{Newton} SOC for the great support in scheduling our observations.
AM acknowledges a financial support from the agreement ASI-INAF n.2017-14-H.0 (PI: T. Belloni, A. De Rosa), the HERMES project by the Italian Space Agency (ASI) n. 2016/13 U.O, the H2020 ERC Consolidator Grant “MAGNESIA” No. 817661 (PI: Rea) and National Spanish grant PGC2018-095512-BI00.

%%%%%%%%%%%%%%%%%%%%%%%%%%%%%%%%%%%%%%%%%%%%%%%%%%



\bibliographystyle{mnras}
\bibliography{bibliography} 
%%%%%%%%%%%%%%%%%%%%%%%%%%%%%%%%%%%%%%%%%%%%%%%%%%

%%%%%%%%%%%%%%%%% APPENDICES %%%%%%%%%%%%%%%%%%%%%

%\appendix


%%%%%%%%%%%%%%%%%%%%%%%%%%%%%%%%%%%%%%%%%%%%%%%%%%

% Don't change these lines
\bsp	% typesetting comment
\label{lastpage}

\end{document}